\RequirePackage{etoolbox}
\csdef{input@path}{%
 {sty/}
 {img/}
}%
\csgdef{bibdir}{bib/}

\documentclass[ba]{imsart}
\pubyear{}
\volume{}
\issue{}

\usepackage{amsfonts,amssymb,graphics,epsfig,verbatim,bm,latexsym,amsmath,url,amsbsy}
\usepackage{multirow}
\usepackage{booktabs}
\usepackage{amsthm}
\usepackage{amsmath}
\usepackage{natbib}
\usepackage{changepage}
\usepackage{float}
\usepackage[colorlinks,citecolor=blue,urlcolor=blue,filecolor=blue,backref=page]{hyperref}
\usepackage{graphicx}
\usepackage{tablefootnote}
\usepackage{threeparttable}
\startlocaldefs
\numberwithin{equation}{section}
\theoremstyle{plain}
\endlocaldefs

\newtheorem{assumption}{Assumption}[section]
\newtheorem{proposition}{Proposition}[section]

\renewcommand{\thesection}{\arabic{section}}

\def\AmSTeX{$\cal A$\kern-.1667em\lower.5ex\hbox{$\cal M$}\kern-.125em
    $\cal S$-\TeX}
\def\BibTeX{{\rm B\kern-.05em{\sc i\kern-.025em b}\kern-.08em
    T\kern-.1667em\lower.7ex\hbox{E}\kern-.125emX}}

\DeclareMathOperator{\argmin}{argmin}
\DeclareMathOperator{\argmax}{argmax}

\DeclareMathOperator{\diag}{diag}
\DeclareMathOperator{\tr}{tr}
\DeclareMathOperator{\VEC}{vec}

\begin{document}

\begin{frontmatter}
\title{Accurate Computation of Marginal Data Densities Using Variational Bayes \thanksref{T1}}
\thankstext{T1}{The authors thank Bill Griffiths for his useful comments on the paper.}

\begin{aug}

\author{\fnms{Gholamreza} \snm{Hajargasht}\thanksref{addr1} and \fnms{Tomasz} \snm{Wo\'zniak}\thanksref{addr2}}

\runtitle{MDD using Variational Bayes}

\ead[label=e1]{rhajargasht@swin.edu.au}
\ead[label=e2]{tomasz.wozniak@unimelb.edu.au}

\address[addr1]{Author's Address: Swinburne Business School, BA Building, 27 John St, Hawthorn Victoria 3122, Australia, email: \printead{e1}}

\address[addr2]{Author's Address: University of Melbourne, Department of Economics, FBE Building, Level 4, 111 Barry St, Carlton, Victoria 3053, Australia, email: \printead{e2}}


\end{aug}

\begin{abstract}
We propose a new marginal data density estimator (MDDE) that uses the variational Bayes posterior density as a weighting density of the reciprocal importance sampling (RIS) MDDE. This computationally convenient estimator is based on variational Bayes posterior densities that are available for many models and requires simulated draws only from the posterior distribution. It provides accurate estimates with a moderate number of posterior draws, has a finite variance, and provides a minimum variance candidate for the class of RIS MDDEs. Its reciprocal is consistent, asymptotically normally distributed, and unbiased. 
These properties are obtained without truncating the weighting density, which is typical for other such estimators. Our proposed estimators outperform many existing MDDEs in terms of bias and numerical standard errors. In particular, our RIS MDDE performs uniformly better than other estimators from this class.
\end{abstract}



\end{frontmatter}

\section{Introduction}\label{sec:introduction}\setcounter{equation}{0} 

\noindent The marginal data density (MDD), also known as marginal likelihood, is  of central importance in Bayesian inference, providing a summary of the evidence contained in the data about a model. Since it is required for the computation of Bayes factors and posterior odds ratios \citep[see e.g.][]{Kass1995}, it is essential for model comparisons, predictive analyses, and Bayesian hypotheses assessment.  

To define the MDD, let $\theta$ denote a vector collecting all of the parameters of a model and $\mathbf{y}$ denote observed data. MDD is the constant normalizing the posterior distribution of parameters given the data which can be obtained by the application of Bayes' rule:
\begin{equation}\label{eq:bayesrule}
p\left(\theta|\mathbf{y}\right) = \frac{p\left(\mathbf{y}|\theta\right)p\left(\theta\right)}{p\left(\mathbf{y}\right)},
\end{equation}
where on the left hand side is the posterior distribution, the numerator on the right hand side of equation (\ref{eq:bayesrule}) is the product of the likelihood function and the prior distribution of the parameters of the model, and the denominator is the MDD. The latter is often obtained through integrating the joint distribution of data and parameters over the $n$-dimensional parameter vector defined on the parameter space $\Theta\subseteq\mathbb{R}^n$:
\begin{equation}\label{eq:MDD}
p\left(\mathbf{y}\right)  = \int_{\theta\in\Theta}p\left(\mathbf{y}|\theta\right)p\left(\theta\right) d\theta.
\end{equation}

The integral in equation (\ref{eq:MDD}) can be computed analytically only for simple models. 
For more complicated models, elaborate numerical integration methods are required \citep[see e.g.,][]{Gelfand1990}. Estimation of MDD using such methods often involves significant costs due to substantial programming and computational requirements. Nevertheless, pursuing numerically precise and computationally feasible estimation techniques is worthwhile because of the importance of MDD applications. Good summaries of various proposals can be found in \cite{Ardia2012} and \cite{chen2012monte} among others.

We propose two new MDD estimators (MDDEs) that outperform existing alternatives in terms of the numerical efficiency while maintaining computational feasibility. The first of these estimators belongs to the class of reciprocal importance sampling (RIS) MDDEs and the second to the class of Bridge Sampling (BS) MDDEs. The former class was proposed by \cite{Gelfand1994} and defined by:
\begin{equation}\label{eq:MDD-ris}
\hat{p}_{RIS}(\mathbf{y}) = \Big(\frac{1}{S}\sum_{s=1}^{S}\frac{h\left(\theta^{(s)}\right)}{p\left(\mathbf{y}|\theta^{(s)}\right)p\left(\theta^{(s)}\right)}\Big)^{-1},
\end{equation}
 where $\left\{ \theta^{(s)} \right\}_{s=1}^{S}$ is a sample drawn from the posterior distribution and $h(\theta)$ is a weighting density. The properties of the weighting function determine the properties of the estimator. For this estimator to have desirable properties, the support of the weighting density must be a subset of the posterior density's support or equivalently the ratio $h(\theta)/p(\mathbf{y}|\theta)p(\theta)$ must be bounded from above \citep[see e.g.][] {Gelfand1994, chen2012monte}. Under this condition, $\hat{p}_{RIS}(\mathbf{y})^{-1}$ is a consistent estimator of $p(\mathbf{y})^{-1}$ \citep[see e.g.][]{Geweke2005}. 
 
An early proposal (harmonic mean MDDE) by \cite{Newton1994} to set $h(\theta)=p(\theta)$ did not satisfy this condition whenever the prior density had fatter tails than the posterior. This fact led researchers to propose weighting densities such that their support is a subset of the posterior density's support. For instance, \cite{Geweke1999} proposed using a truncated $n$-dimensional normal density given by:
\begin{equation}\label{eq:geweke}
h(\theta) = (1-\alpha)^{-1}(2\pi)^{-\frac{n}{2}}\det(\bar{V}_{\theta})^{-\frac{1}{2}}
\exp\Big( -\frac{1}{2} (\theta-\bar\theta)'\bar{V}_{\theta}^{-1}(\theta-\bar\theta) \Big)I_{\left( (\theta-\bar\theta)'\bar\Omega^{-1}(\theta-\bar\theta) \leq \chi^2_{\alpha}(n) \right)},
\end{equation}
where $I_{(\cdot)}$ is the indicator function taking the value of one if the condition in the parentheses holds and zero otherwise, and $\chi^2_{\alpha}(n)$ is the $100(1-\alpha)$th percentile of the chi-square distribution with $n$ degrees of freedom. \cite{Geweke1999} set $\bar\theta$ and $\bar{V}_{\theta}$ to the posterior mean and covariance of the parameters respectively, and $\alpha$ to 0.05, which usually guarantees that the posterior distribution dominates the weighting function.

\cite{Sims2008} used a truncated elliptical density as the weighting function. Let $\hat{\theta}$ denote the mode of the posterior distribution and define a scaling matrix by $\hat\Omega = S^{-1}\sum_{s=1}^{S}(\theta^{(S)} - \hat\theta)(\theta^{(S)} - \hat\theta)'$. Define a set $\Theta_L=\left\{ \theta: p(\mathbf{y}|\theta)p(\theta)>L \right\}$ and denote by $(1-\alpha_L)$ a fraction of the posterior draws belonging to set $\Theta_L$. $L$ is chosen so that $(1-\alpha_L)\approx 0.9$. Then the weighting density proposed by \cite{Sims2008} is given by:
\begin{equation}\label{eq:swz}
h(\theta) = \Gamma\left(\frac{n}{2}\right)\frac{\nu}{b^{\nu}-a^{\nu}}(1-\alpha_L)^{-1}(2\pi)^{-\frac{n}{2}}\big|\det\Big(\hat\Omega^{\frac{1}{2}}\Big)\big|^{-1}
\Big( (\theta-\hat\theta)'\hat\Omega^{-1}(\theta-\hat\theta) \Big)^{\frac{\nu-n}{2}}
I_{\left( \theta\in\Theta_L \right)},
\end{equation}
where parameters $a$, $b$, and $\nu$ are determined as in \cite{Sims2008}. Compared to  \citeauthor{Geweke1999}'s weighting function, the proposal of \citeauthor{Sims2008} involves more computational costs, mainly associated with finding the mode of the posterior distribution. However, it offers improved numerical efficiency especially for models with posterior distributions of nonstandard shapes. These methods require some arbitrary choices regarding the truncation and can require many draws from the posterior distribution to achieve reasonable numerical precision. 

Another solution has been to change the space of the weighting function through analytical integration \citep[see, e.g.,][]{FuentesAlberoMelosi2013}. However, this suggestion is applicable only if appropriate analytical integration is possible for a given model. Finally, \cite{Lenk2009} proposed correcting the harmonic mean MDDE by using the ratio of the prior and posterior probabilities on an appropriately defined subset of the support of posterior density. This solution leads to desirable properties of the estimator such as having a~finite variance.\footnote{See also \cite{Pajor2016} who applies the same strategy to the arithmetic mean MDDE. Another paper attempting to correct harmonic mean estimator is \cite{raftery2006estimating}.} 

Our proposal for the RIS MDDE is to use the output from variational Bayes (VB) estimation as the weighting function. The objective of VB estimation is to approximate the posterior density with another density that is tractable. The VB approximate posterior density is optimal in the sense that it minimizes the KL divergence from this density to the posterior density. We show that a density constructed in this way, under some mild regularity conditions, is dominated by the posterior density, i.e., it has a~support that is a subset of the support of the posterior density and, thus, is a good candidate for a weighting function for the RIS MDDE.

We also show that the aforementioned optimality of the VB approximate density translates into optimality of the resulting MDDE. It leads to an RIS MDDE that is minimum variance within the class of RIS MDDEs. This finding is further confirmed by our simulations where we show that our estimator has the smallest numerical standard errors within the class of RIS MDDEs and also it performs well when compared to some popular alternative MDDEs from other classes of estimators. Finally, we show that our new MDDE has the desirable properties that were also established by \cite{Geweke2005} and \cite{Fruhwirth-Schnatter2004} for other RIS MDDEs. In particular, we show that our estimator is consistent, asymptotically normally distributed with a finite variance, and unbiased. Our results are obtained under less restrictive assumptions compared to those by \cite{Geweke2005}. In particular, we relax the requirement of truncating the weighting density and  prove that the posterior density always dominates the VB approximate posterior density. This last feature, inherent in the VB candidate, leads to the desired properties of our new RIS MDDE. 

The focus of this paper is on the RIS MDDE because it can be easily computed given that a sample from the posterior distribution is available.\footnote{Most of known methods of obtaining a sample from the posterior distribution is suitable for our purposes including direct sampling, Markov Chain Monte Carlo methods, Importance Sampling, etc.} In contrast to many other MDDEs, the RIS MDDE does not require any other simulations. This is a~desirable feature especially in those cases where the cost of additional computations is particularly high, for instance, due to a large dimension of the parameter space. The last point is well illustrated in recent developments in macroeconometrics where hypothesis testing for large nonlinear dynamic models is essential. \cite{FuentesAlberoMelosi2013} and \cite{Droumaguet2017} emphasize the feasibility of MDD estimation using an adjusted version of the estimator by \cite{Geweke1999,Geweke2005} whereas \cite{Sims2008} propose a~new weighting function that assures the numerical stability of the RIS MDDE in models with nonstandard posterior distributions. Other recent applications of the RIS MDDE in the field literature include \cite{Waggoner2012a}, \cite{Castelnuovo2014}, \cite{Hubrich2015}, \cite{Burda2015}, \cite{Carriero2016}, and \cite{Hajargasht2018}.

Although our main focus is on the properties and performance of the RIS MDDE, We also investigate the usefulness of using the VB approximate density as a weighting function for the BS MDDE proposed by \cite{Meng1996} \citep[see also][for a recent review]{Gronau2017}. This estimator is obtained through an iterative procedure initialized at some starting value for $\hat{p}_{BS}^{(0)}(\mathbf{y})$ and the following recursion:
\begin{equation}\label{eq:MDD-bs}
\hat{p}_{BS}^{(t)}(\mathbf{y}) = \hat{p}_{BS}^{(t-1)}(\mathbf{y}) \frac{ \frac{1}{O}\sum_{o=1}^{O}\frac{\hat{p}( \theta^{(o)}|\mathbf{y} ) }{ Og( \theta^{(s)} ) + S\hat{p}( \theta^{(o)}|\mathbf{y} )} }{
\frac{1}{S}\sum_{s=1}^{S}\frac{g( \theta^{(s)} )}{Og( \theta^{(s)} ) + S\hat{p}( \theta^{(s)}|\mathbf{y} )}
},
\end{equation}
where $\{ \theta^{(o)} \}_{o=1}^{O}$ is an i.i.d. sample drawn from the weighting density~$g(\theta)$, $\{ \theta^{(s)} \}_{s=1}^{S}$ is an Markov Chain Monte Carlo (MCMC) sample drawn from the posterior, and $\hat{p}(\theta|\mathbf{y}) = p(\mathbf{y}|\theta)p(\theta)/\hat{p}_{BS}^{(t-1)}(\mathbf{y})$. In practice, the iterative procedure presented in equation (\ref{eq:MDD-bs}) requires few steps, usually around ten, until convergence is achieved. One advantage of the BS MDDE is that, irrespective of whether the weighting density is dominated by the posterior density or not, it has finite variance \citep[see][]{Fruhwirth-Schnatter2004}. Consequently, it is easy to see that the desired properties hold if the weighting function $g(\theta)$ is set to the VB approximate posterior density. 

Another appealing feature of our estimators is that VB posterior densities are available for many models through a vast number of papers that provide VB estimation for different type of models. Additionally, recent software packages, Infer.Net \citep{Minka2012}, STAN \citep{Kucukelbir2015} and R package rstan by \cite{rstan}, provide automatic VB posterior density estimation. Thus, the range of models to which our method can be easily applied is enormous.

The paper is organized as follows: In section \ref{sec:vb}, we present preliminaries on VB methods to the extent required for our application to the MDDE and establish the dominance property. Section~\ref{sec:vbMDDenew} introduces the new RIS MDDE, states its properties and briefly reviews some of the other MDDEs that are evaluated in our simulation experiments. In section \ref{sec:empirical} and Appendix C, simulation experiments are used to study the numerical accuracy of the MDDE in three applications: vector autoregressions, stochastic frontier and longitudinal Poisson models. We have chosen vector autoregressions since it is possible to calculate the exact MDD for this model under a conjugate prior assumption making it useful for comparisons. The Stochastic Frontier models are also considered as an example of a nonlinear model with latent variable.Appendix A provides proofs of the propositions, Appendix B provides the computational details of VB estimation of the considered models and Appendix C reports the comparison of numerical accuracy of the MDDEs for a Poisson Model estimated for epileptic seizure counts data.

\section{Variational Bayes Methods}\label{sec:vb}\setcounter{equation}{0}

\noindent In this section we introduce a general setting of VB estimation and establish a property of the resulting approximate posterior density that is essential for MDD estimation.

\subsection{Variational Bayes Estimation}\label{ssec:vb}

\noindent VB is a deterministic alternative to MCMC; it was developed as a Bayesian method of inference in machine learning \citep[see e.g.][]{attias2000variational,jordan1999introduction}. Recent surveys of VB include \cite{wainwright2008graphical}, \citeauthor{Bishop2006}~(\citeyear{Bishop2006}, Chapter~10), and \cite{Murphy2012}, amongst others.  
Some asymptotic properties of the VB estimator for a wide range of models can be found in \cite{Wang2017} and the references therein. In this section, we follow the notation and the basic setting of VB method in \cite{Ormerod2010} and \cite{Blei2017}.

The objective of Bayesian inference is the characterization of the posterior distribution of the parameters of a model given data. In VB the idea is to approximate the posterior density with a density $q(\theta)$ which is of a tractable form. The optimal approximate posterior density, denoted by $q^*\left(\theta\right)$, is obtained by minimizing the Kullback-Leibler divergence of the density $q\left(\theta\right)$ from the true posterior density, $p\left(\theta|\mathbf{y}\right)$:
\begin{equation}\label{eq:minkl}
q^*(\theta) = \underset{q \in \mathcal{Q}}{\argmin} KL[q(\theta)||p(\theta|\mathbf{y})] 
= \min_{q\in\mathcal{Q}} \int_{\theta\in\Theta}q(\theta)\ln\Big( \frac{q(\theta)}{p(\theta|\mathbf{y})} \Big)d\theta,
\end{equation}
where $\mathcal{Q}$ is a family of densities.

In order to make the VB method appealing, the approximate density $q^*\left(\theta\right)$ should be of an analytically tractable form. This can often be achieved by imposing simplifying assumptions on $q$. One such assumption commonly used is that the approximate density of $\theta$ can be expressed as a product of factorized densities for some partition $\{ \theta_1, \dots, \theta_M\}$ of vector $\theta$:
\begin{equation}\label{eq:vbassumption}
q(\theta) = \prod_{m=1}^{M}q_m(\theta_m).
\end{equation}
This factorized form corresponds to an approximation framework developed in physics known as \emph{mean-field theory}. Thus, this form of variational approximation is often referred to as mean-field VB. 

The assumption given in equation (\ref{eq:vbassumption}) introduces conditional stochastic independence between sub-vectors of parameters $\theta_m$, for $m\in\{1,\dots, M\}$, given data $\mathbf{y}$. Using results from calculus of variations and some simple algebraic manipulations it is shown that the optimal factorized densities $q_m^*(\theta)$ can be obtained from the following iterative procedure \citep[see][]{Ormerod2010}: Initialize $q_{2}(\theta_2), \dots, q_{M}(\theta_M)$, and iterate:
\begin{align*}
q_1(\theta_1) &\leftarrow \frac{\exp( E_{-\theta_1}\ln( p(\mathbf{y},\theta) ) )}{\int \exp( E_{-\theta_1}\ln( p(\mathbf{y},\theta) ) ) d\theta_1}\\
&\vdots\\
q_M(\theta_M) &\leftarrow \frac{\exp( E_{-\theta_M}\ln( p(\mathbf{y},\theta) ) )}{\int \exp( E_{-\theta_M}\ln( p(\mathbf{y},\theta) ) ) d\theta_M},
\end{align*}
until the decrease in the value of the KL divergence, $KL[q\left(\theta\right)||p\left(\theta|\mathbf{y}\right)]$, between two subsequent iterations is negligible. $E_{-\theta_m}$ denotes expectation with respect to the density $\prod_{j\neq m}q_j(\theta_j)$.

For many econometric models, imposing assumption (\ref{eq:vbassumption}) leads to factorized distributions belonging to known families of parametric distributions. For such cases, let the approximate density $q(\theta)$ be parameterized by a vector of hyper-parameters $\lambda$, defined on a set of their admissible values $\Lambda$, and denoted by  $q(\theta|\lambda)$. The optimal $q^*(\theta|\lambda)$ is defined by the optimal values of its hyper-parameters $\lambda^*$ obtained by minimizing the Kullback-Leibler divergence of the approximate density, $q\left(\theta|\lambda\right)$, from the true posterior density, $p(\theta|\mathbf{y})$ over the hyper-parameters $\lambda\in\Lambda$. 
Then the iterative optimization procedure for the distributions described above transforms into a deterministic iterative procedure for the hyper-parameters or moments of these distributions which is known as the \emph{coordinate ascent algorithm}. The convergence of this deterministic algorithm is often obtained in a small number of steps, which makes VB estimation much faster than estimation based on MCMC \citep[see][for some recent analysis of the properties of the coordinate ascent algorithm for the mean-field VB]{Zhang2017}.

\subsection{The Convergence Criterion}\label{ssec:convergence}

\noindent Note that the constant normalizing the kernel of the posterior distribution is unknown and, in consequence, the value of the KL divergence is unknown as well. To facilitate the computations, the natural logarithm of $p(\mathbf{y})$ is expressed as a sum of a value defined by $\ln MDD_{VBLB}(\lambda) = E_q \ln p(\mathbf{y},\theta) - E_q \ln q(\theta|\lambda)$ and KL divergence measure, $KL[q(\theta|\lambda)||p(\theta|\mathbf{y})]$ \citep[see e.g.][p. 142]{Ormerod2010}:
\begin{equation}\label{eq:vbml}
\ln p(\mathbf{y}) = \ln MDD_{VBLB}(\lambda) + KL[q(\theta|\lambda)||p(\theta|\mathbf{y})].
\end{equation}
A closed-form solution for $\ln MDD_{VBLB}(\lambda)$ can be derived for many models with little effort. Moreover, since the KL divergence measure is nonnegative, and is equal to zero only when $p(\theta|\mathbf{y})=q^*(\theta|\lambda)$, $\ln MDD_{VBLB}(\lambda)$ provides a lower bound to $\ln MDD$. The problem of minimizing $KL[q(\theta|\lambda)||p(\theta|\mathbf{y})]$ stated in equation (\ref{eq:minkl}) is equivalent to the problem of maximizing $\ln MDD_{VBLB}(\lambda)$ with respect to $\lambda$:
\begin{equation}\label{eq:maxvblb}
\lambda^* = \underset{\lambda\in\Lambda }{\argmax} \ln MDD_{VBLB}(\lambda)
= \underset{\lambda \in\Lambda}{\argmax} [E_q \ln p(\mathbf{y}|\theta)p(\theta) - E_q \ln q(\theta|\lambda)].
\end{equation}
Consequently, to monitor the convergence of the coordinate ascent algorithm $\ln MDD_{VBLB}(\lambda)$ is used instead of KL distance as the objective function .

\subsection{The Dominance Property}\label{sec:dominanceprop}

\noindent In this work we adopt the following assumptions in relation to the dominance property:

\begin{assumption}\label{as:densities}
For any given data $\mathbf{y}$, the likelihood function is bounded from above for all values of the parameter vector $\theta$ in the parameter space. 
\end{assumption}
\begin{assumption}\label{as:2}
The prior density $p(\theta)$ is proper and bounded from above.
\end{assumption}
\begin{assumption}\label{as:3}
The VB approximate posterior density $q^*(\theta)$ is proper and bounded from above.
\end{assumption}
\begin{assumption}\label{as:4}
The VB approximate posterior density  $q^*(\theta)$ is continuous and differentiable. 
\end{assumption}
The first two assumptions are uncontroversial and guarantee the existence of the posterior distribution with a finite MDD. The other two conditions define the VB approximate posterior density in quite general terms.

Define sets $\Omega_p=\{\theta: p(\theta|\mathbf{y})>0\}$ and $\Omega_q=\{\theta: q^*(\theta)>0\}$ being the supports of the posterior density and  of the approximate density respectively. Denote by $\Omega^c$ the complement of set $\Omega$. Under Assumptions \ref{as:densities}--\ref{as:4} we show that the posterior density dominates the VB approximate posterior density $q^*(\theta)$.
\begin{proposition}\label{cl:dominance}
 Let Assumptions \ref{as:densities}--\ref{as:4} hold and the posterior density $p(\theta|\mathbf{y})$ be continuous and differentiable on its support $\Omega_p$. If a solution $q^*(\theta)$ to the minimisation problem from equation \eqref{eq:minkl} exists, then the posterior density dominates the VB approximate posterior density, i.e., $p(\mathbf{y}|\theta) p(\theta)=0\Rightarrow q^*(\theta)=0$, almost everywhere.
\end{proposition}
This proposition states that the solution to the optimization problem such as the one given in equation (\ref{eq:minkl}) has to meet the dominance property. Otherwise, $q^*(\theta)$ cannot be a solution because then the value of the KL distance goes to infinity. This property seems to be well-known, however, we have not found its formal proof. Furthermore, it has some desirable consequences that have been discussed, for instance, in \citeauthor{Murphy2012}~(\citeyear{Murphy2012}, Chapter~21).

\section{The New MDD Estimator}\label{sec:vbMDDenew}\setcounter{equation}{0}

\noindent Our new RIS MDDE is defined by assigning the VB approximate posterior density as the weighting density $h(\theta)=q^*(\theta)$ in equation (\ref{eq:MDD-ris}). This MDDE is given by:
\begin{equation}\label{eq:MDDrisvb}
\hat p_{RIS.VB}(\mathbf{y}) = \Big( \frac{1}{S}\sum_{s=1}^{S}\frac{q^*(\theta^{(s)})}{p(\mathbf{y}|\theta^{(s)})p(\theta^{(s)})} \Big)^{-1}.
\end{equation}
The fact that the VB approximate posterior density is dominated by the posterior density is essential for the properties of the new RIS MDDE. We present them below.

\subsection{Properties of the New MDD Estimator}\label{sec:vbMDDe}

\noindent In this section, we consider the properties of the RIS MDDE, given in equation (\ref{eq:MDDrisvb}), with the VB approximate posterior density used as the weighting function. We show that this estimator is indeed theoretically well-justified. All of the proofs are given in the Appendix.

The first two propositions set the objects of interest that are referred to in the propositions and proofs. Proposition \ref{prop:MDDinv} suggests that the estimator given in equation (\ref{eq:MDDrisvb}) is an unbiased estimator for the inverse normalizing constant, whereas proposition \ref{prop:var} shows that $Var_p\big(\frac{q^*}{p}\big)$ can be written as an expectation which is further used in our discussion of the MDDE variance properties.

\begin{proposition}\label{prop:MDDinv}
If Assumptions \ref{as:densities}--\ref{as:4} hold, then:
$$\int_{\Omega_p}^{}\frac{q^*(\theta)}{p(\mathbf{y}|\theta)p(\theta)}p(\theta|\mathbf{y})d\theta=\frac{1}{p(\mathbf{y})}.$$
\end{proposition}
\begin{proposition}\label{prop:var}
If Assumptions \ref{as:densities}--\ref{as:4} hold, then $Var_p\big(\frac{q^*}{p}\big)=E_q\big[\frac{q^*}{p}\big]-1$.
\end{proposition}
\noindent Note that in order for the variance in the proposition above to be finite it is required that:
\begin{equation*}
E_q\left[\frac{q^*}{p}-1\right]=\int_{\Omega_q} \left[\frac{q^{*}(\theta)}{p(\theta|\mathbf{y})}-1\right]q^*(\theta)d\theta<\infty,
\end{equation*}
on which we argue below.

Subsequently, we require that a sample from the posterior distribution is uniformly ergodic which we state in an assumption below.
\begin{assumption}\label{as:ergodic}
The draws $\left\{ \theta^{(s)} \right\}_{s=1}^{S}$ are from a uniformly ergodic process with unique invariant density $p(\theta|\mathbf{y})$ having support $\Omega_p\subseteq\mathbb{R}^n$.
\end{assumption}

\noindent Given this and other assumptions, proposition \ref{prop:pris} states that the RIS-MDDE is consistent and asymptotically normally distributed.

\begin{proposition}\label{prop:pris}
Suppose that Assumptions \ref{as:densities}--\ref{as:4} and \ref{as:ergodic} hold and that $Var_p \Big(\frac{q}{p} \Big)=E_q\left[\frac{q}{p}-1\right]<\infty$. Then:
\begin{description}
\item[(i)] $\sqrt{S}(\hat{p}_{RIS.VB}(\mathbf{y})^{-1}-p(\mathbf{y})^{-1}) \overset{d}{\rightarrow} N\Big(0,\frac{\rho_d(0)}{p(\mathbf{y})^2}Var_p \Big(\frac{q}{p} \Big)\Big),$ 
\item[(ii)] $E[ \hat{p}_{RIS.VB}(\mathbf{y})^{-1} ] = p(\mathbf{y})^{-1}$,
\end{description}
where $\overset{d}{\rightarrow}$ denotes convergence in distribution at the limit where $S\rightarrow\infty$, $\frac{q}{p}$ denotes the ratio $q^*(\theta)/p(\theta|\mathbf{y})$, and $\rho_h(0)$ is the normalized spectral density  of the process $d_s=\frac{q^*(\theta^{(s)})}{p(\mathbf{y}|\theta^{(s)})p(\theta^{(s)})}$ at frequency zero.
\end{proposition}

\noindent The result from part(i) of the proposition above is not unknown for the class of the RIS estimators proposed by \cite{Gelfand1994}. One important assumption in the proof stated above is finitness of variance $Var_p \Big(\frac{q}{p} \Big)$. In some of the the existing approaches, such as those proposed by \cite{Geweke1999,Geweke2005} and \cite{Sims2008}, an arbitrary truncation of the weighting density, $h(\theta)$, assures that this variance is finite. However, the cost of the truncation is that these approaches require a large number of MCMC draws to provide an accurate MDD estimate, as we show in section \ref{sec:empirical}. The truncation of the weighting function is not required with VB approximate posterior density as the weighting function. There are good theoretical reasons to expect $q^*(\theta)$ as the weighting function to yield a finite variance for RIS MDDE. In fact, the variance is expected to be near minimum within the family of considered approximate densities $\mathcal{Q}$. Proposition~\ref{prop:minvar} states that a variation of VB provides a minimum variance estimator within the class of the Reciprocal Importance Sampling MDDEs. 
\begin{proposition}\label{prop:minvar}
The RIS MDDE with the VB posterior as the weighting density, under a $\chi^2$ divergence criterion, is minimum variance among the class of RIS MDDEs with weighting densities from $\mathcal{Q}$.
\end{proposition}

\noindent Note that where KL is minimized the ratio of $q^*/p$ should be close to one for all $\theta$ and, as a result, $E_q ln\left[\frac{q^*}{p}\right]$ should be well approximated with $E_q\left[\frac{q^*}{p}-1\right]$, the expectation from proposition \ref{prop:var}. Therefore, one should expect a small MDDE variance for optimal $q$ based on a KL divergence as well. This feature manifests itself in an improved numerical efficiency of our estimator compared to these alternatives which we clearly document in section \ref{sec:empirical}.  

Both of the results in proposition \ref{prop:pris} are stated in terms of the reciprocal of the estimator following similar such results in \cite{Geweke1999,Geweke2005} and \cite{Fruhwirth-Schnatter2006} for the RIS MDDE. The relevance of such statements is motivated by a common practice of reporting the logarithm of MDDEs. We report the negative of logarithm of the inverse of our unbiased estimator i.e. $-\ln \left(\hat{p}_{RIS.VB}(\mathbf{y})^{-1}\right)$. Note that $\hat{p}_{RIS.VB}(\mathbf{y})^{-1}$  is the unbiased estimator of the reciprocal MDD. Therefore, reporting its negative logarithm is similar to reporting the logarithm of an unbiased MDDE such as the importance sampling MDDE. 

We conclude with a remark about the effects of potential problems with the convergence of the coordinate ascent algorithm used to compute the VB approximate posterior density, $q^*(\theta)$, on the properties of our estimator. In a general case, the algorithm provides a local solution to the optimization problem that might not be global optimum. However, even a local solution exhibits the dominance property which is required for our RIS MDDE to remain consistent, unbiased, asymptotically normal, and of finite variance. 

\subsection{Comparison with Existing MDD Estimators}\label{sec:comparison}

\noindent Conceptually, the idea of using the VB approximate posterior density as the weighting function under an RIS framework is close to the idea of employing entropy methods to obtain the optimal importance density for the importance sampling (IS) MDDE proposed by \cite{ChanEisenstat2015}. The IS MDDE itself was introduced by \cite{VanDijk1978} and \cite{Geweke1989} and is given by:
\begin{equation}\label{eq:MDD-is}
\hat{p}_{IS}(\mathbf{y}) = \frac{1}{R}\sum_{r=1}^{R}\frac{p(\mathbf{y}|\theta^{(r)})p(\theta^{(r)})}{f(\theta^{(r)})},
\end{equation}
where $\{ \theta^{(r)} \}_{r=1}^{R}$ is an i.i.d. sample drawn from the importance density $f(\theta)$. In order to have a good estimator, the importance density should be a close approximation to the posterior distribution. Its choice is a nontrivial task that is discussed below. From the point of view of the MDD estimation, it is essential that the importance density dominates the posterior distribution. This property, in many references, is described as $f$ having thicker tails than the posterior distribution. \cite{ChanEisenstat2015} show that if this is the case then the IS MDDE is consistent, unbiased, and asymptotically normally distributed, as well as having a finite variance. Note that the construction of an appropriate importance density is the most challenging task associated with this estimator.

\cite{ChanEisenstat2015} propose using an importance density that is derived from cross-entropy methods. Their method requires defining a parametric family of distributions parametrized by hyper-parameters $\kappa$. A distribution belonging to such a family, denoted by $\tilde{f}(\theta|\kappa)$, is used to approximate the posterior distribution well. The \emph{optimal} hyper-parameters, $\kappa^*$, are obtained by minimizing the Kullback-Leibler divergence of the the true posterior density, $p(\theta|\mathbf{y})$, from the approximate density, $\tilde{f}(\theta|\kappa)$, over the admissible set of hyper-parameters' values $\mathcal{K}$:
\begin{equation}\label{eq:minkl-crossentropy}
\kappa^* = \underset{\kappa\in\mathcal{K}}{\argmin} KL[p(\theta|\mathbf{y}) || \tilde{f}(\theta|\kappa)]
= \min_{\kappa\in\mathcal{K}} \int_{\theta\in\Theta}p(\theta|\mathbf{y})\ln\Big[ \frac{p(\theta|\mathbf{y})}{\tilde{f}(\theta|\kappa)} \Big]d\theta.
\end{equation}
In practice, finding the optimal hyper-parameters is equivalent to solving the following problem:
\begin{equation}\label{eq:maxkappa}
\kappa^* = \underset{\kappa\in\mathcal{K}}{\argmax} \frac{1}{S}\sum_{s=1}^{S}\ln \tilde{f}( \theta^{(s)}|\kappa ),
\end{equation}
where a sample of $S$ draws from the posterior distribution is used. Subsequently, the importance density is set to $q(\theta) = \tilde{f}(\theta| \kappa^* )$ and used in the estimation of the MDD according to equation (\ref{eq:MDD-is}). \cite{ChanEisenstat2015} show that under the assumption that $\tilde{f}(\theta| \kappa^* )$ dominates the posterior distribution and therefore this MDDE is consistent and unbiased.

Note that while our approximate posterior density minimizes the \emph{reverse} KL distance between this density and the true posterior density (see equation \ref{eq:minkl}), the optimal importance density is obtained by minimizing the \emph{forward} KL distance between the true posterior and the importance density (see equation \ref{eq:minkl-crossentropy}). 

By the same argument that we used to prove proposition \ref{cl:dominance}, it can be shown that the optimal importance density dominates the posterior density
, a property assumed by \cite{ChanEisenstat2015}. Consequently, the entropy method delivers a weighting function that is suitable for the IS MDDE, while the VB method delivers one that is appropriate for the RIS MDDE.

Nevertheless, there are significant differences between the two approaches to MDD estimation. The method by \citeauthor{ChanEisenstat2015} does not require sampling from the posterior distribution but rather uses independent sampling from $f(\theta)$. Note that in many cases the location and precision of $f(\theta)$ are determined using a sample from the posterior distribution and, thus, sampling from importance density constitutes additional computational cost. Further, this method requires some additional simulations for the optimization process and, more importantly, making some arbitrary choices regarding the functional form of the importance density. Our estimator requires only a sample from the posterior density and it does not require any other simulations. Moreover, in the case of using the mean-field VB the approximate posterior density, $q(\theta)$ is determined by the assumptions regarding the likelihood function and prior distributions and therefore free of arbitrary choices. The hyper-parameters of the approximate densities are determined through the coordinate ascent algorithm that is fast and automatic. Therefore, our proposition requires relatively little computational effort.

Another related method for choosing the weighting function within the IS MDDE framework, studied in detail by \cite{Perrakis2014}, is by using the product of marginal posterior densities (PMPD) \citep[see also][]{Fruhwirth-Schnatter2004} given by:
\begin{equation}\label{eq:productofmarginal}
\prod_{m=1}^{M}p(\theta_m| \mathbf{y} ),
\end{equation}
where the ordinates of the marginal densities are usually computed using the Rao- Blackwellization. This approach can be thought as the dual of our RIS MDDE proposal in the sense that, in mean-field VB, the approximate posterior density is obtained by taking expectation of the posterior kernel with respect to $q(\theta)$ while under the product of marginal densities framework, the weighting density is obtained by taking expectation with respect to $p(\theta|\mathbf{y})$. In comparison, (i) The VB approximate posterior density is dominated by the true posterior density which makes it a good candidate for RIS MDDE while the opposite is true for the product of marginals which makes it a good candidate for IS MDDE; (ii) It has also been shown by \cite{Botev2013}, that the product of marginals is the best importance density, in the sense of minimizing forward Kullback-Leibler divergence from the true posterior density and, therefore, minimizing the MDDE variance, among all product form importance densities. As we showed, the same is true for VB in the context of reverse Kullback-Leibler divergence and reciprocal importance sampling; (iii) An advantage of our proposal is that in many cases the VB approximate posterior density can be obtained analytically or by simple optimizations while the product of marginals requires both simulation and estimation of the marginal posterior densities. We estimate MDDs using the product of marginals whenever it is feasible in our simulation comparisons.

\section{Numerical Accuracy of the MDD Estimators for Two Models}\label{sec:empirical}\setcounter{equation}{0}




\noindent We study the performance of the new MDDEs by applying them to three classes of models: vector autoregressions with two alternative prior specifications, stochastic frontier models with two alternative specifications of the inefficiency effect. A similar study of the numerical accuracy of the MDDEs the longitudinal Poisson model is reported in the Appendix. 

For each model we compute benchmark values of the logarithm of the MDD, i.e. $\ln\hat{p}(\mathbf{y})$. If possible, an exact benchmark is computed by analytical integration. Otherwise, we use the estimators by \cite{Chib1995} as a~benchmark. We also report the VB lower-bound of the logarithm of the MDDE, denoted below by $\ln MDD_{VBLB}$ and its upper bound counterpart using a method proposed by \cite{ji2010bounded}. All of the MDDEs included in our comparison of numerical accuracy and the corresponding references are given in the Table \ref{tab:MDDes}.

\begin{table}[t]
\raggedright
\caption{MDD estimators used in comparisons of numerical accuracy}
\begin{center}
\small
\begin{tabular}{lll}
\toprule
\multicolumn{3}{c}{\textbf{RIS MDDEs by \cite{Gelfand1994} based on equation \eqref{eq:MDD-ris}}}\\[1ex]
 & $h(\theta)$  & reference \\ 
 \midrule 
 VB & $q^*(\theta)$  & the current paper\\
 Geweke & see eq. \eqref{eq:geweke}   & \cite{Geweke1999,Geweke2005}\\
 Sims et al. & see eq. \eqref{eq:swz}  & \cite{Sims2008}\\
 PMPD & see eq. \eqref{eq:productofmarginal}  & the current paper\\
 Newton, Raftery & $p(\theta)$ & \cite{Newton1994} \\
 Lenk & $p(\theta)$    & \cite{Lenk2009} \\
 \midrule
 \multicolumn{3}{c}{\textbf{BS MDDEs by \cite{Meng1996} based on equation \eqref{eq:MDD-bs}}}\\[1ex]
 & $g(\theta)$   &  \\
 \midrule
 VB & $q^*(\theta)$   & the current paper \\
 PMPD & see eq. \eqref{eq:productofmarginal}  & \cite{Fruhwirth-Schnatter2004} \\
 Normal & $\mathcal{N}(\bar{\theta},\bar{V}_{\theta})$  & \cite{DiCiccio1997} \\

Warp3 & $\mathcal{N}(\mathbf{0}_{n\times1},I_n)$ & \cite{meng2002warp} \\ 
\midrule
 \multicolumn{3}{c}{\textbf{IS MDDEs by \cite{Geweke1989} based on equation \eqref{eq:MDD-is}}}\\[1ex]
 & $f(\theta)$   &  \\ 
 \midrule
 VB & $q^*(\theta)$  & the current paper \\
 PMPD & see eq. \eqref{eq:productofmarginal}  & \cite{Perrakis2014}\\
Chan, Eisenstat &  & \cite{ChanEisenstat2015}\\
\midrule
 \multicolumn{3}{c}{\textbf{Other MDDEs}}\\[1ex]
 \midrule
 exact & & \cite{Karlsson2013}\\
 Chib & & \cite{Chib1995}\\
\bottomrule
\end{tabular}\\
\end{center}
\label{tab:MDDes}
\footnotesize  \renewcommand{\baselineskip}{11pt}
Note: Explanations of acronyms: 
PMPD - product of marginal posterior densities,
VB - variational Bayes.
The computations for the Warp3 BS MDDE were performed using the R package by \cite{Gronau2018}.
\end{table}

For each model and data set, we compute values of the log-MDDs a hundred times using independent MCMC simulations. Each of these simulations include 10,000 draws from the stationary posterior distribution. Subsequently, we use these values to compute their average value for each of the MDDEs reported in tables \ref{tab:nsevars} and \ref{tab:nsesfm} and two criteria for the assessment of the numerical accuracy of alternative MDDEs, namely, the numerical standard error (NSE), and the fraction of simulated MDD values that fall in the interval between the VB lower and upper bounds for the $\ln MDD$ mentioned above, denoted in tables by "\%~within". The NSE is defined as the sample standard deviation for the MCMC sample of simulated values of $\ln\hat{p}(\theta)$. Finally, all of the VB estimation details and the formulae for the VB lower bounds of the $\ln MDDs$, $\ln MDD_{VBLB}$, for the models presented in this section can be found in the Appendix.

\subsection{Vector Autoregressive Model}

\noindent Consider the VAR($p$) model:
\begin{equation}\label{eq:varp}
y_t = a_0 + A_1y_{t-1} + \dots + A_py_{t-p} + \epsilon_t, \text{ and } \epsilon_t \sim i.i.d.\mathcal{N}_{N}(\mathbf{0}_{N\times 1}, \Sigma),
\end{equation}
for $t\in\{1,\dots,T\}$, and $T$ denoting the sample size, $p$ is the lag order of the model, $y_t$ is an $N$-vector of observations at time $t$, $a_0$ is a vector of constant terms and $A_i$ for $i\in\{1,\dots,p\}$ are matrices of autoregressive coefficients. $\epsilon_t$ is an  error term that is conditionally normally distributed given past observations with the mean being a~vector of zeros and the covariance matrix $\Sigma$.

Define a $T\times N$ matrix $Y = (y_1, \dots, y_T)'$ collecting all the observations of vector $y_t$. Define a vector $x_t = (1, y_{t-1}', \dots , y_{t-p}')'$, a matrix $X = (x_1, \dots, x_T)'$, and the matrix of coefficients $A = ( a_0, A_1, \dots, A_p )'$. Also, let matrix $E = (\epsilon_1, \dots, \epsilon_T)'$ collect all the error terms. Then the VAR process from equation (\ref{eq:varp}) can be written as:
\begin{equation}\label{eq:varpmatricnorm}
Y = XA + E,\qquad E \sim \mathcal{MN}_{T\times N}(\mathbf{0}_{T\times N}, \Sigma, I_T),
\end{equation}
where $E$ follows a matrix-variate normal distribution \citep[see e.g.][]{Wozniak2016} with the mean set to a matrix of zeros and the covariance matrix for a vectorized vector $e = \text{vec}(E')$ given by $\Sigma\otimes I_T$, where $\otimes$ denotes the Kronecker product. 

For illustrative purposes, we consider two alternative prior distributions for VAR models: a natural-conjugate distribution for which the posterior distribution, VB approximate distribution, and the MDD can be derived analytically, and an independent prior distribution that results in iterative procedures for the estimation of the same quantities.

\subsubsection{Prior and posterior analysis with VB estimation}

\noindent For convenience, we introduce another matrix representation. Define vectors $y = \text{vec}(Y')$ and $\alpha = \text{vec}(A)$. Then, the model from equation (\ref{eq:varp}) can also be written as:
\begin{equation}\label{eq:varpnorm}
y = (I_N\otimes X) \alpha + e,\quad e \sim \mathcal{N}_{TN}(\mathbf{0}, \Sigma\otimes I_T).
\end{equation}
Below, we present the details of VB estimation; while the details of Gibbs sampling can be found in  \cite{Karlsson2013}.

\paragraph{Normal-Wishart Conjugate Prior} 
For VAR models with normally distributed error terms the natural-conjugate prior distribution is given in the following normal-Wishart form:
\begin{equation*}
p(A,\Sigma) = p(A|\Sigma)p(\Sigma^{-1}),\quad
A|\Sigma \sim\mathcal{MN}_{K\times N}(\underline{A}, \Sigma,\underline{V}  ), \quad
\Sigma^{-1}\sim\mathcal{W}( \underline{S}^{-1}, \underline{\nu} ).
\end{equation*}
where $\underline{A}$ is a $(1+pN)\times N$ matrix of prior means, $\underline{V}$ is a positive-definite matrix of order $(1+pN)$, $\underline{S}$ is an $N\times N$ positive-definite scale matrix of the Wishart prior distribution, and $\underline{\nu}$ denotes its degrees of freedom parameter.

Consider firstly the assumption that VB does not factorize the parameters into sub-groups, but instead a joint distribution $q_{A\Sigma}(A,\Sigma^{-1} )$ is derived for $A$ and $\Sigma^{-1}$ \emph{en bloc}. In such a case, VB inference is equivalent to exact posterior analysis in which the joint posterior distribution of $(A,\Sigma^{-1})$ is of Normal-Wishart form \citep[see e.g.][]{Wozniak2016}:
\begin{equation}
q(A,\Sigma^{-1} ) \equiv p(A,\Sigma^{-1}|\mathbf{y} ) = \mathcal{NW}(\overline{A},\overline{V}, \overline{S}^{-1}, \overline{\nu} ).
\end{equation}
where:
\begin{align*}
\overline{A} &= \overline{V}(\underline{V}^{-1}\underline{A} + X^{'}Y ), \\
\overline{V} &= ( \underline{V}^{-1} + X^{'}X )^{-1},\\
\overline{S} &= (Y-X\overline{A})^{'}(Y-X\overline{A}) + \underline{S} + (\overline{A} - \underline{A})^{'}\underline{V}(\overline{A}-\underline{A}), \\
\overline{\nu} &= T+\underline{\nu}.
\end{align*}
Importantly, the marginal posterior distribution of matrix $A$ is given by the following matrix-variate $t$-distribution \citep[see the definition of this matrix-variate distribution in][Appendix~A]{Bauwens1999}
\begin{equation}\label{eq:marginalA}
p(A|\mathbf{y} ) = \mathcal{M}t(\overline{A}, \overline{V}, \overline{S}^{-1}, \overline{\nu} ).
\end{equation}
Additionally, the MDD can be computed as an ordinate of the matrix-variate $t$-distribution at the data vector $Y$ \citep[see][]{Karlsson2013}:
\begin{equation}
Y \sim \mathcal{M}t(X\underline{A}, (I_T + X\underline{V}X^{'})^{-1}, \underline{S}, \underline{\nu} ).
\end{equation}

Consider now VB inference that is derived under the assumption that the VB approximate posterior distribution is factorized into marginal distributions for $A$ and $\Sigma^{-1}$:
\begin{equation}
q(A,\Sigma^{-1}) = q_A(A )q_{\Sigma^{-1}}(\Sigma^{-1}).
\end{equation}
The optimal VB approximate posterior distributions are given as closed-form formulae with the matrix-variate normal distribution for $A$ and the Wishart distribution for $\Sigma^{-1}$:
\begin{align*}
q_{A}^{*}(A) &=\mathcal{MN}_{N.K}( \overline{A}^{*}, (\overline{\nu}^{*}-N-1)^{-1}\overline{S}^{*},\overline{V}^{*} ), \\
q_{\Sigma}^{*}(\Sigma^{-1}) &=\mathcal{W}( \overline{S}^{*-1}, \overline{\nu}^{*} ),
\end{align*}
where:
\begin{align*}
\overline{A}^{*} & = \overline{V}^{*}( \underline{V}^{-1}\underline{A} + X^{'}Y ),\\
\overline{V}^{*} & = ( \underline{V}^{-1} + X^{'}X )^{-1},\\
\overline{\nu}^{*} &= T + 1+pN + \underline{\nu},\\
\overline{S}^{*} &= \frac{\overline{\nu}^{*}}{\overline{\nu}^{*}- 1-pN} (( Y-X\overline{A}^{*} )^{'}( Y-X\overline{A}^{*} ) + \underline{S} + ( \overline{A}^{*} - \underline{A} )^{'}\underline{V}^{-1}( \overline{A}^{*} - \underline{A} )).
\end{align*}

This example of a linear Gaussian model illustrates well the working of our RIS MDDE with the VB approximate posterior density used as the weighting function. The exact posterior density is of the normal-Wishart form that implies the marginal distribution for matrices $A$ and $\Sigma$ given by the matrix-variate $t$-distribution, as in equation \eqref{eq:marginalA}, and the Wishart distribution respectively. The corresponding marginal VB posterior distributions are matrix-variate normal and Wishart. It is easy to see in this case, that the exact posterior distribution dominates its approximate counterpart as the latter has thinner tails. This feature is general as shown in our proposition \ref{cl:dominance} and implies desirable properties of our new RIS MDD estimator.


\paragraph{Normal-Wishart Independent Prior}
This prior distribution presumes that the auto- regressive parameters $\alpha$ and the precision matrix $\Sigma^{-1}$ are \emph{a priori} independent:
\begin{equation*}
p(\alpha,\Sigma^{-1}) = p(\alpha)p(\Sigma^{-1}), \quad
\alpha \sim\mathcal{N}_{NK}(\underline{\alpha}, \underline{\underline{V}}), \quad
\Sigma^{-1}\sim\mathcal{W}( \underline{S}^{-1}, \underline{\nu}),
\end{equation*}
where $\underline{\alpha}$ is an $N(1+pN)$-vector of prior means and $\underline{\underline{V}}$ is a positive-definite covariance matrix of order $N(1+pN)$.

The optimal VB approximate posterior distributions are the multivariate normal distribution for $\alpha$ and the Wishart distribution for $\Sigma^{-1}$:
\begin{equation}
q_{\alpha}^{*}(\alpha) =\mathcal{N}( \overline{\alpha}^{*}, \overline{V}^{*} ),\text{ and }
q_{\Sigma}^{*}(\Sigma^{-1}) =\mathcal{W}( \overline{S}^{*-1}, \overline{\nu}^{*} ),
\end{equation}
where the optimal values of the parameters determining these distributions are obtained using an iterative procedure that we describe in the Appendix. 


\subsubsection{Comparison of the MDD estimators}

To study the performance of the new MDDEs for the VAR models we use the same data set as \cite{Giannone2015} with the same variable transformations. The time series consist of seven quarterly variables that span the period starting in the first quarter of 1959 and finishing in the last quarter of 2008, which gives 200 observations. The seven variables that we consider are real GDP, GDP deflator, federal funds rate, real consumption, real investment, hours worked and real compensation per hour. We estimated a model with four autoregressive lags ($p=4$). We base our estimations on $S=10,000$ draws from the posterior distribution.

\begin{table}[t!]
\raggedright
\caption{Numerical accuracy of the MDDEs for VAR models}
\begin{center}
\footnotesize
\begin{tabular}{lcccccc}
\toprule
\multicolumn{7}{c}{\textbf{Natural-conjugate prior distribution}}\\
&\multicolumn{6}{l}{\textit{Benchmark values}}\\
&	 exact & VB$_{LB}$ & VB$_{UB}$  &&&\\
$\ln\hat{p}(\mathbf{y})$&	2910.1& 2908.2& 2912.3&&&\\
NSE&	-&	-&	0.025&&&\\
\midrule
&\multicolumn{6}{l}{\textit{RIS MDDEs}}\\
$h(\theta)$ & VB & Geweke & Sims et al. & PMPD & Newton, Raftery & Lenk \\
$\ln\hat{p}(\mathbf{y})$&	\textbf{2910.1}&	2907.5&	2907.9 &	\textbf{2910.1}	&		3168.7&	2905.6\\
NSE&	0.062&	0.254&	0.284&	0.094&	3.251&	2.543\\
\% within & 100&	0&	6 &	100	&	0&	10 \\
\midrule
&\multicolumn{3}{l}{\textit{BS MDDEs}}&\multicolumn{3}{l}{\textit{IS MDDEs}}\\
$g(\theta)$ & VB & PMPD & Normal  &   VB & PMPD & Chan, Eisenstat\\
$\ln\hat{p}(\mathbf{y})$&	\textbf{2910.1}&	\textbf{2910.1}&	\textbf{2908.8} & \textbf{2910.1} & \textbf{2910.1}&\textbf{2910.1}\\
NSE&	0.017&	0.015&	0.024&	0.101& 0.0888&0.177\\
\% within & 100 & 100 & 100& 100 & 100& 100 \\

\midrule
\multicolumn{7}{c}{\textbf{Independent prior distribution}}\\
&\multicolumn{6}{l}{\textit{Benchmark values}}\\
&	 Chib & VB$_{LB}$ & VB$_{UB}$ &&& \\
$\ln\hat{p}(\mathbf{y})$&	\textbf{2882.1} & 2880.4 & 2884.1 &&&\\	
NSE&	0.111 &-& 0.022&&&	\\
\midrule
&\multicolumn{6}{l}{\textit{RIS MDDEs}}\\
$h(\theta)$ & VB & Geweke & Sims et al. &  PMPD& Newton, Raftery & Lenk \\
$\ln\hat{p}(\mathbf{y})$&	\textbf{2882.1} & 2879.4 & 2879.8 & \textbf{2882.1} & 3167.5 & 2877.7 \\
NSE&	0.047 & 0.173 & 0.216 &	0.099 & 3.209 & 2.412 \\
\% within & 100 & 0 & 0& 100 & 0 & 10\\
\midrule
&\multicolumn{3}{l}{\textit{BS MDDEs}}&\multicolumn{3}{l}{\textit{IS MDDEs}}\\
$g(\theta)$ & VB & PMPD & Normal  & VB & PMPD &Chan, Eisenstat\\
$\ln\hat{p}(\mathbf{y})$&	\textbf{2882.1} & \textbf{2882.1} & \textbf{2880.8} & \textbf{2882.1} & \textbf{2882.3}&\textbf{2882.1} \\
NSE&	0.017 & 0.017 & 0.021 &	0.170 & 0.087 & 0.211 \\
\% within & 100 & 100 & 100  & 100 & 100 & 100 \\
\bottomrule
\end{tabular}
\end{center}
\label{tab:nsevars}
\footnotesize
 \renewcommand{\baselineskip}{11pt}
Note: The table reports the average of the natural logarithms of MDDEs, the standard errors, and the fraction of these simulated MDDEs included in the interval between the VB lower and upper bounds reported in the top row of each panel. These values are computed via simulations based on 100 independent repetitions. The values in bold face denote the MDDE values that lay within the interval bounds. See table \ref{tab:MDDes} for detailed descriptions of the reported MDDEs. 
\end{table}
In the top panel of Table \ref{tab:nsevars} we report the results for assessment of the numerical accuracy of the MDDEs for the VAR model with the normal-Wishart conjugate prior distribution. For this model the benchmark value of the MDD can be computed exactly. Note that it lies in the middle of the interval between the VB lower and upper bounds for the MDD.
Our estimator is the best estimator within the family of the RIS MDDEs. It has the lowest NSE and 100 percent of simulated MDDEs fall within the VB bounds. The value of the NSE for our estimator is 30 percent lower than the second best specification, which is, the RIS MDDE with the product of marginal posterior densities used as the weighting function. The estimators of \cite{Geweke1999,Geweke2005} and \cite{Sims2008} perform quite badly in terms of both the NSE and the fraction of simulated MDDE values within the VB bounds. In our simulations, $S$ had to be increased over fivefold to make these estimators any competitive in these terms. Moreover, despite the fact that the VB approximate density is dominated by the exact posterior density, the VB IS MDDE still performs reasonably. Its NSE is just slightly higher than that of the best estimator in IS MDDE class that uses PMPD as the weighting function and it outperforms the estimator by \cite{ChanEisenstat2015}. Finally,  in the case of the BS MDDEs, VB and the PMPD candidates perform very well although PMPD is slightly better than our estimator.

These findings are confirmed by Figure 1 that illustrates the values of the logarithm of the simulated MDDEs for 100 repetitions of the estimation for the best four MDDEs. According to Figure 1, the best estimators in terms of precision are BS PMPD and BS VB followed by RIS VB and IS PMPD. We do not spot any significant bias in these estimators as their values oscillate around the benchmark. 

The results for the VARs with independent prior distributions for parameters matrices $A$ and $\Sigma$ are reported in the bottom panel of table \ref{tab:nsevars}. Here, the MDDEs based on the VB approximate posterior density are the best in terms of the NSE relative to their counterparts within the RIS MDDE and BS MDDE classes. Note that our BS MDDE performs better than two benchmark specifications by \cite{Chib1995} and \cite{Fruhwirth-Schnatter2006}. 

\begin{figure}[H]
\raggedright
\caption{Comparison of MDDEs simulation outcomes  for Vector Autoregressions}
\begin{center}
\begin{tabular}{c}
\textbf{Natural-conjugate prior distribution}\\
\includegraphics[trim=1cm 0.5cm 1cm 1.5cm, scale=0.35]{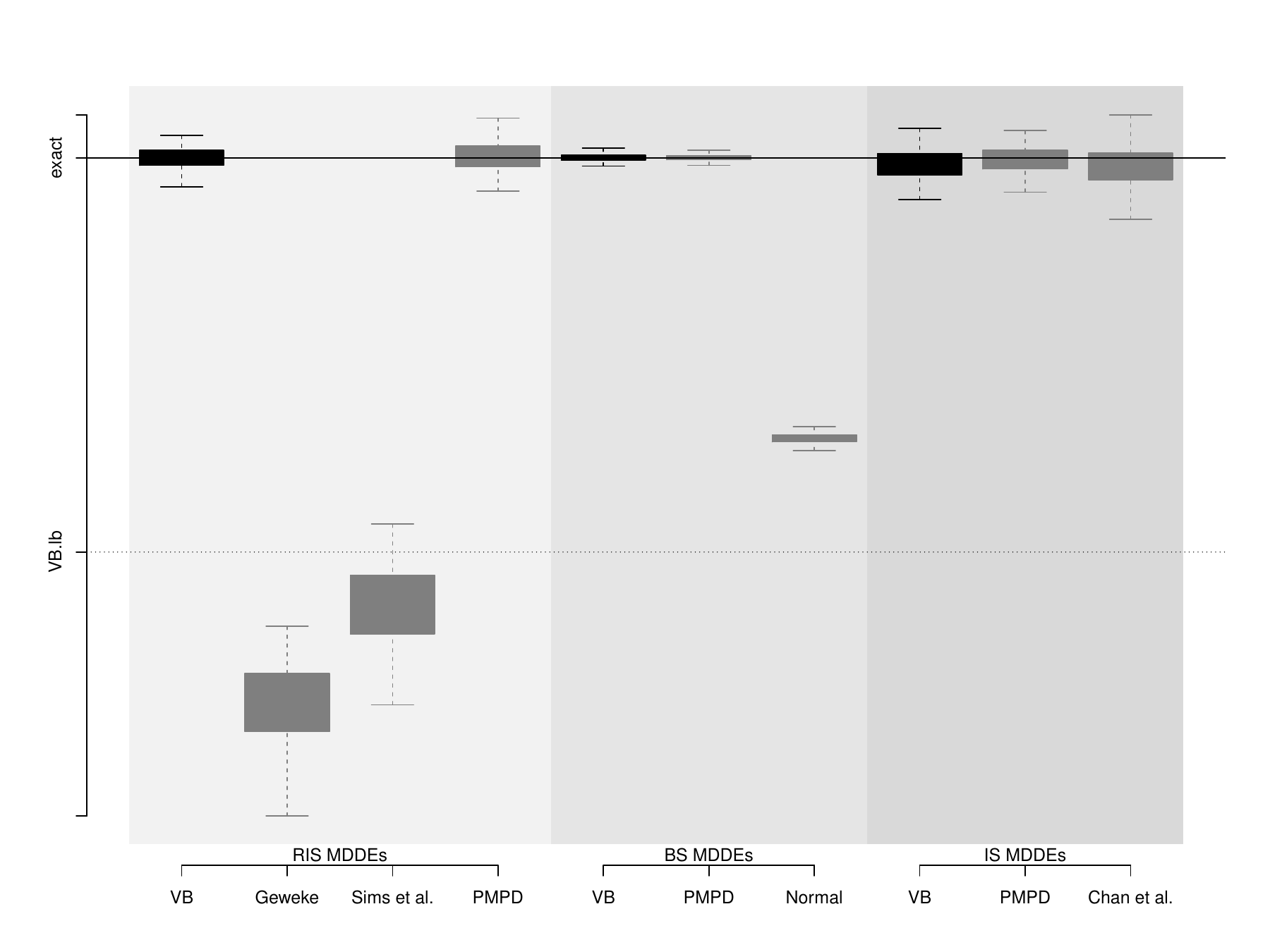}\\
\textbf{Independent prior distribution}\\
\includegraphics[trim=1cm 1cm 1cm 1.5cm, scale=0.35]{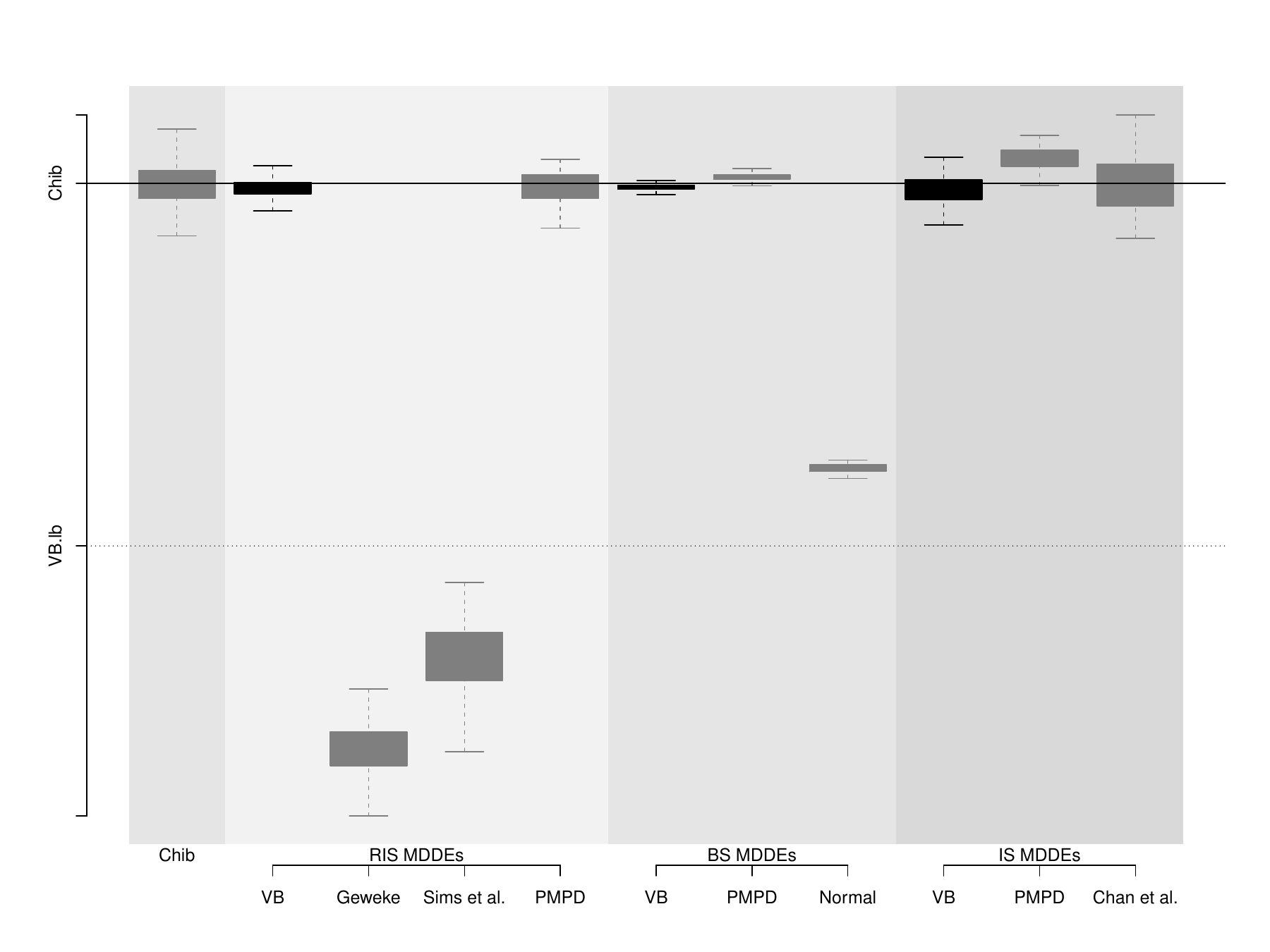}
\end{tabular}
\label{fig:MDD-var-nc}
\end{center}
\footnotesize
 \renewcommand{\baselineskip}{11pt}
Note: The figure reports box plots of the natural logarithms of the MDDEs for 100 independent repetitions of the estimation process. Black box of MDDEs with VB approximate posterior distribution as a weighting function are black. The horizontal line in the first plot corresponds to the exact value of the natural logarithms of the MDD whereas in the second one -- to a chosen benchmark estimated value of the MDD. See table~\ref{tab:MDDes} for detailed descriptions of the reported MDDEs. Two RIS MDDEs with the largest NSEs (Newton \& Raftery and Lenk) are omitted from the plots.
\end{figure}

These findings are confirmed by figure \ref{fig:MDD-var-nc} that reports the box plots of the logarithm of the simulated MDDEs. Black box plots in this figure correspond to our new MDDEs with VB approximate posterior distributions used as weighting functions. Note that these estimators perform very well compared to other estimators within particular class. They are not biased and the precision of the estimation is often greater than that of the competing estimators. It is also worth emphasizing that the VB lower bound for $\ln MDDE$ seems to be an efficient tool to discriminate against estimators with large biases such as those for most of the RIS MDDEs except for two of them .

\subsection{Stochastic Frontier Model} 

\noindent Consider a stochastic frontier (SF) model with panel data that can be written as:
\begin{equation}\label{eq:sfm}
y_{it}= x_{it}\beta \pm u_i +v_{it},
\end{equation}
where $i=1,\ldots ,N$ indexes firms and $t=1,\dots ,T$ indexes time, $y_{it}$ is the dependent variable (either firms' output or cost), $x_{it}$ is $k$-dimensional row vector of regressors, $x_{it}\beta$ is the log of the frontier production or cost function, $\beta$ is a $k$-vector of unknown parameters, $u_i$ is a non-negative random error reflecting the inefficiency of firm $i$, and $v_{it}$ is an i.i.d. normally distributed error term with mean zero and constant variance $\sigma^2$, denoted by $v_{it}\sim\mathcal{N}( 0,\sigma^2 )$. The negative sign before $u_i$ is for a production frontier model and the plus sign is for the cost frontier case. Finally, let an $N\times1$ vector $u=(u_1,, \dots, u_N)'$ collect all of the inefficiency terms.

The Bayesian approach to estimation of stochastic frontier models using Monte Carlo methods has been described in \cite{Broeck1994} and \cite{Koop2001} among others. In this section, we consider stochastic frontier models with two alternative distributions for inefficiency term, namely, exponential and gamma. Further details about the mean-field VB estimation for these models can be found in \cite{Hajargasht2018}. For  MCMC estimation of theses models see \cite{Koop2001} and \cite{Tsionas2000}. Using an exponential distribution leads to optimal approximate posterior densities of standard forms, while this is not the case for the gamma-distributed inefficiency case where some numerical integration is required to obtain VB posterior distributions.

\paragraph{Exponential inefficiency}
In this case we consider the stochastic frontier model in equation (\ref{eq:sfm}) with $u_i$ following an exponential distribution with parameter $\lambda$, i.e., $p(u_i|\lambda)=\lambda \exp ( -\lambda u_i )$.

\paragraph{Gamma inefficiency}
The alternative inefficiency error distribution that we consider is gamma, $u_i\sim \mathcal{G}(\theta ,\lambda )$. This case is interesting from the point of view of VB estimation because it results in some approximate posterior densities that are not of a known form. In these cases, it is necessary to use numerical integration to obtain some of the required moments.

\subsubsection{Prior and posterior analysis with VB estimation}

\paragraph{Exponential inefficiency}
We assume the following prior distributions:
\begin{equation}\label{eq:sfmeprior}
p(\beta)\sim \mathcal{N}(\underline{\beta},\underline{V}_{\beta}), \qquad
\sigma^{-2}\sim\mathcal{G}(\underline{A}_{\sigma}, \underline{B}_{\sigma} ),\qquad
\lambda\sim\mathcal{G}(\underline{A}_{\lambda},\underline{B}_{\lambda} )
\end{equation}
where $\mathcal{G}(\underline{A},\underline{B})$ is the gamma density function as defined in \cite{Hajargasht2018}. The prior distributions for $\sigma^2$ and $\lambda$ are standard choices in the Bayesian stochastic frontier literature. 

To facilitate mean-field VB estimation we need an appropriately factorized approximation to the posterior distribution. We consider the following:
\begin{equation}
q(\beta, \sigma^{-2}, u, \lambda)=q_{\beta}(\beta) q_{\sigma^{-2}}\left(\sigma^{-2}\right) q_{\lambda}(\lambda)q_u(u) 
\end{equation}
The optimal VB approximate posterior distributions turn out to be:
\begin{equation*}
q_{\beta}^{*}(\beta)= \mathcal{N}(\overline{\beta}^{*},\overline{V}_{\beta}^{*} ), \quad
q_{\sigma^{-2}}^{*}(\sigma^{-2})= \mathcal{G}(\overline{A}_{\sigma}^{*}, \overline{B}_{\sigma}^{*}), \quad
q_{\lambda}^{*}(\lambda)= \mathcal{G}(\overline{A}_{\lambda}^{*}, \overline{B}_{\lambda}^{*} ), \quad
q_u^{*}(u_i)= \mathcal{TN}(\overline{\mu}_i^{*},\overline{\upsilon}^{2*} ),
\end{equation*}
where $\mathcal{TN}(\cdot ,\cdot )$ denotes the normal density function truncated to positive values of $u_i$. The optimal hyper-parameters characterizing these distributions are computed by the coordinate ascent algorithm that we describe in the online supplement.

\paragraph{Gamma inefficiency}
Following \cite{GriffinSteel2007}, we use the following prior distributions: 
\begin{equation}
\lambda|\theta \sim\mathcal{G}\left( \theta ,\underline{B}_{\lambda} \right), \qquad
\theta^{-1}\sim\mathcal{G}\left(\underline{A}_{\theta},\underline{B}_{\theta}\right), 
\end{equation}
while for $\beta$ and $\sigma^2$ we assume the same distribution as presented in equation (\ref{eq:sfmeprior}) for the exponential efficiency case. 

By applying a suitable factorization, it can be shown that the VB optimal densities are given by:
\begin{equation}
q\left(\beta, \sigma, \lambda, \theta, u\right) = 
q_{\beta}(\beta)q_{\sigma}\left(\sigma^{-2}\right)q_{\lambda}(\lambda)q_{\theta}(\theta)q_u\left(u\right),
\end{equation}
The first three of these distributions have functional forms of standard distributions:
\begin{equation*}
q_{\beta}^{*}(\beta) = \mathcal{N}(\overline{\beta}^{*},\overline{V}_{\beta}^{*} ),\quad
q_{\sigma}^{*}(\sigma^{-2}) = \mathcal{G}(\overline{A}_{\sigma}^{*},\overline{B}_{\sigma}^{*} ),\quad
q_{\lambda}^{*}(\lambda) = \mathcal{G}(\overline{A}_{\lambda}^{*},\overline{B}_{\lambda}^{*} ),
\end{equation*}
while the pdf of the approximate posterior distribution for $u$, $q_u^{*}(u_i)$, has a nonstandard form while the one for  $\theta$, $q_{\theta}^{*}(\theta)$, is known up to its normalizing constant. The details of these two distributions as well as on the hyper-parameters of all of the VB approximate posterior distributions specified above are given in the online supplement.

\begin{table}[t!]
\raggedright
\caption{Numerical accuracy of the MDDE for Stochastic Frontier Models}
\begin{center}\small
\begin{tabular}{lccccc}
\toprule
\multicolumn{6}{c}{\textbf{Exponential inefficiency}}\\
&\multicolumn{3}{l}{\textit{Benchmark values}}&&\\
& Chib & VB$_{LB}$ & VB$_{UB}$ && \\
$\ln\hat{p}(\mathbf{y})$&	\textbf{-117.1}& -118.7& -111.4 &&\\
NSE&	0.026 &-& 0.302 && \\
\% within & 100 &  &&&\\
\midrule
&\multicolumn{3}{l}{\textit{RIS MDDEs}}&&\\
$h(\theta)$ & VB & VB CDL & Geweke &  Newton, Raftery & Lenk \\
$\ln\hat{p}(\mathbf{y})$& \textbf{-117.1}& \textbf{-117.1}& \textbf{-114.2}&  -61.6& -122.2\\
NSE&	0.017& 0.041&  0.288&	1.692& 1.629\\
\% within & 100 & 100 & 100&0&2\\
\midrule
\multicolumn{4}{c}{\textit{BS MDDEs}} &&\textit{IS MDDE}\\
$f(\theta)$ & VB & Normal & Warp3&& VB\\
$\ln\hat{p}(\mathbf{y})$&\textbf{-117.1}&  \textbf{-117.1}&\textbf{-117.1}&& \textbf{-117.3}\\
NSE&	 0.021 & 0.059 &0.062&&0.17\\
\% within &   100 & 100 & 100&&100 \\

\midrule
\multicolumn{6}{c}{\textbf{Gamma inefficiency}}\\
&\multicolumn{5}{l}{\textit{Benchmark values}}\\
&  VB$_{LB}$ & VB$_{UB}$ & &&\\
$\ln\hat{p}(\mathbf{y})$& -119.9 &-84.2 & &&\\
NSE&	 & 2.613&& &\\
\midrule
&\multicolumn{5}{l}{\textit{RIS MDDEs}}\\
$h(\theta)$ & VB & Geweke & Newton, Raftery & Lenk&\\
$\ln\hat{p}(\mathbf{y})$& \textbf{-116.3} & \textbf{-114.5} & -62.3 &  -121.5 & \\
NSE&	0.198 & 0.275 & 1.665&	1.823&\\
\% within & 100 & 100 & 0 &  14&\\
\midrule
&\multicolumn{3}{l}{\textit{BS MDDEs}}&&\textit{IS MDDE}\\
$f(\theta)$ & VB & Normal & Warp3& &VB\\		
$\ln\hat{p}(\mathbf{y})$&\textbf{-116.0} & \textbf{-117.2} & \textbf{-117.3} && \textbf{-116.5}\\
NSE&	 0.074 & 0.144 & 0.116 && 0.489\\
\% within &   100 & 100 & 100&&100\\
\bottomrule
\end{tabular}
\end{center}\footnotesize
 \renewcommand{\baselineskip}{11pt}
\label{tab:nsesfm}
Note: The note to Table \ref{tab:nsevars} applies. Additionally, estimators denoted by \emph{VB CDL} involve the complete data likelihood function in the computations in contrast to estimators denoted by \emph{VB} that use the likelihood function with the inefficiency factors integrated out analytically \citep[see][]{ChanGrant2015}.
\end{table}

\subsubsection{Comparison of the MDD estimators}

We illustrate the performance of various MDDEs for the SF models using the data set collected by the International Rice Research Institute and used in such studies as \cite{Coelli2005} and \cite{Hajargasht2018}. We use a panel of 43 Philippine rice farms observed over 4 years from 1990 to 1994. The SF model is a production function with the dependent variable being the logarithm of output, and the regressors being land, hired labour, amount of fertilizer, and other inputs, where all of these values are taken in logarithms. We base our estimations on the final sample of 10,000 draws from the posterior distribution.

In the top panel of Table \ref{tab:nsesfm} we report the results for the assessment of the numerical accuracy of the MDDEs for the SF model with the exponential inefficiency factor. For this model, a full Gibbs sampler can be derived which determines a class of MDDEs that can be applied. As a benchmark, the estimator proposed by \cite{Chib1995} is computed. Note that its value lies much closer to the VB lower bound and is more distant from the upper bound. This is likely due to the fact that the upper bound requires numerical integration over the parameter space as well as over the space of the inefficiency factor. The high dimension of the integral usually leads to inferior precision of estimation. This to some extent is also present when comparing the NSEs for our two RIS MDDEs based on VB approximate posterior density. The estimator denoted by VB uses the likelihood function in the numerator of our RIS MDDE with the inefficiency factor integrated out analytically, while the one denoted by VB CDL is based on the complete data likelihood in which numerical integration has been used. We reach a similar conclusion as \cite{Fruhwirth-Schnatter2008} and \cite{ChanGrant2015} that employment of the complete data likelihood function decreases the numerical efficiency of MDD estimation. However, in our applications, the loss is not as severe as in these papers. The NSE increases around two times but this estimator remains competitive in our comparison.

The main finding from this analysis is that both of our MDDEs based on the VB approximate posterior density as the weighting function, i.e. our RIS MDDE and BS MDDE, have the lowest NSEs. They are also lower than that of the benchmark MDDE of \cite{Chib1995} and nearly fifteen times lower than that of the RIS MDDE proposed by \cite{Geweke1999,Geweke2005}. 

\begin{figure}[H]
\raggedright
\begin{center}
\caption{Comparison of MDDEs simulation outcomes for Stochastic Frontier Models}
\begin{tabular}{c}
\textbf{Exponential inefficiency}\\
\includegraphics[trim=1cm 1cm 1cm 1cm, scale=0.35]{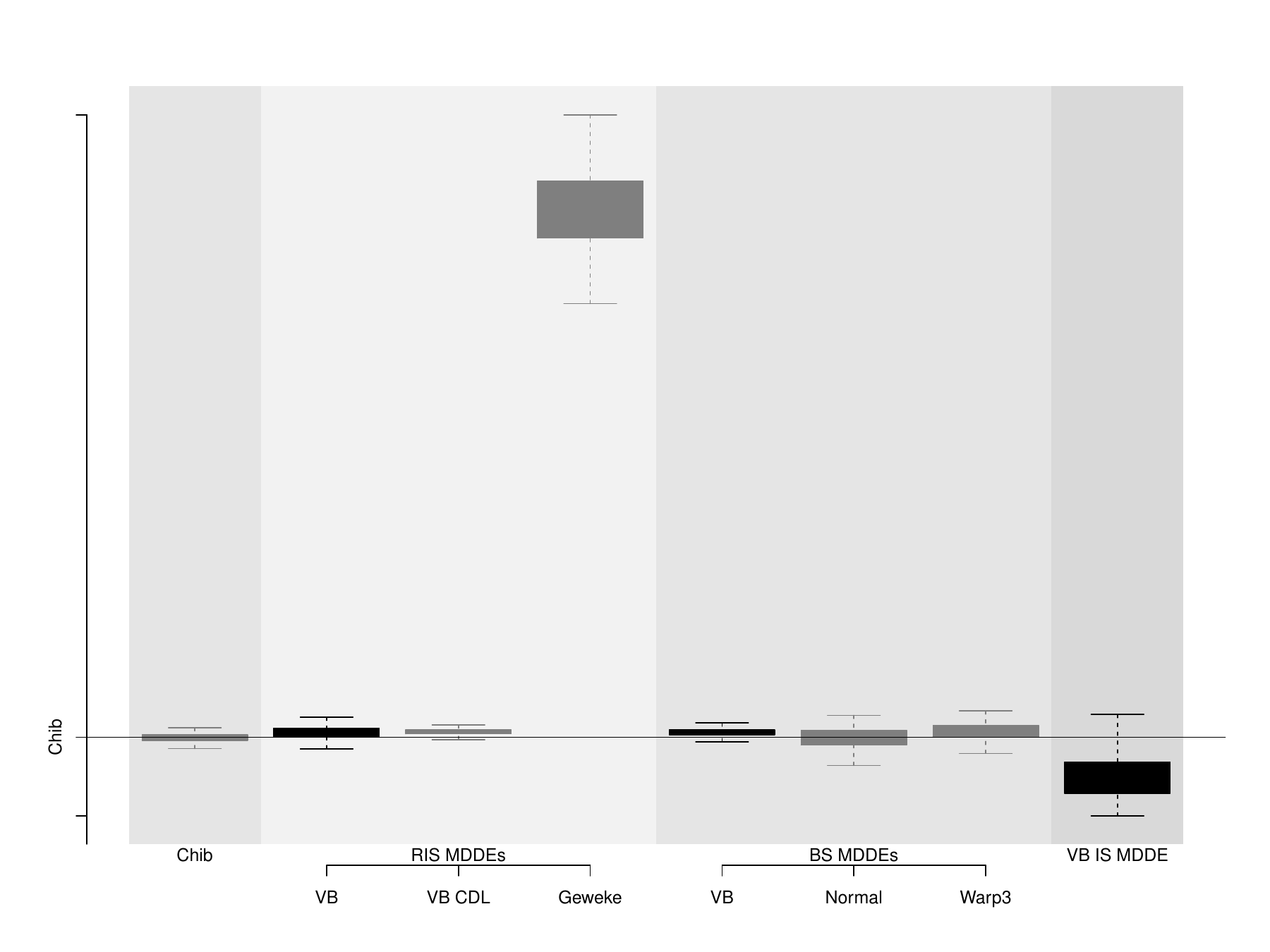}\\
\textbf{Gamma inefficiency}\\
\includegraphics[trim=1cm 1.5cm 1cm 1cm, scale=0.35]{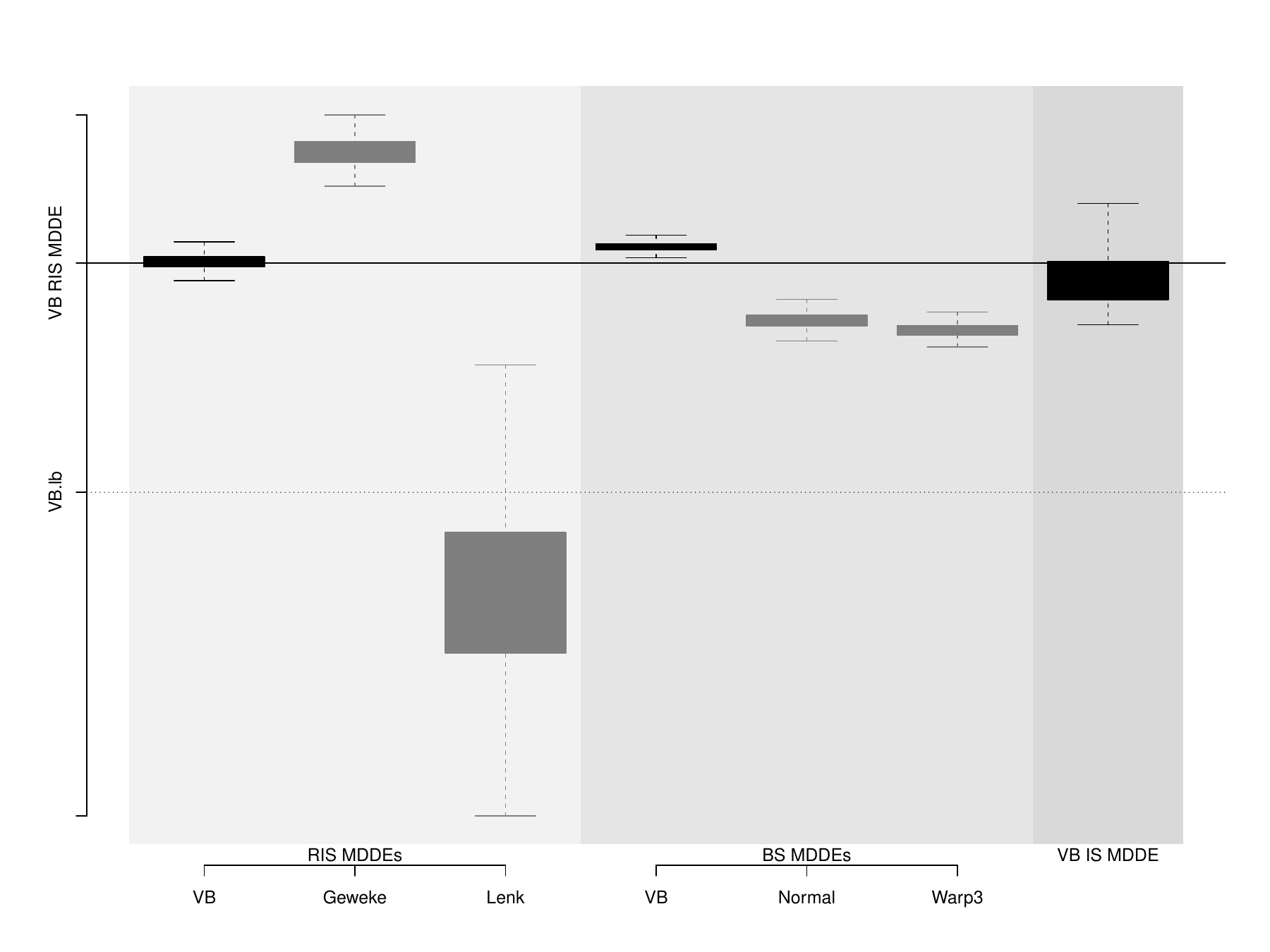}
\end{tabular}
\label{fig:MDD-sf}
\end{center}
\small
\footnotesize
Note: The figure reports box plots of the natural logarithms of the MDDEs for 100 independent repetitions of the estimation process. Black box of MDDEs with VB approximate posterior distribution as a weighting function are black. The horizontal line in the first plot corresponds to the value of Chib's MDDE whereas in the second one -- to the estimated value of the RIS VB MDDE. See Table \ref{tab:MDDes} for detailed descriptions of the reported MDDEs. Two RIS MDDEs with the largest NSEs (Newton \& Raftery and Lenk) are omitted from the plots.
\end{figure}

These findings are confirmed by figure \ref{fig:MDD-sf} that reports the box plots of the logarithm of the simulated MDDEs for 100 repetitions of the estimation for the MDDEs. According to figure \ref{fig:MDD-sf}, VB BS and VB RIS MDDEs with integrated inefficiencies have a similar performance in terms of NSE and bias while Chib's estimates are generally slightly smaller but more varied. Bridge sampling with a "Normal" candidate is much more varied in this case.
The lower panel of Table \ref{tab:nsesfm} includes the results for the SF model with gamma inefficiency. This model is estimated by a Metropolis-Hastings within Gibbs algorithm as only some of the full conditional posterior distributions can be sampled from directly. An appealing feature of the RIS MDDE is that it can be easily applied even in such cases. From the Table we read that our both of our new MDDEs are quite precise and have lower NSEs than their respective counterparts within the RIS MDDE and BS MDDE classes. The lower panel of figure \ref{fig:MDD-sf}  again displays the box plots of the logarithm of the simulated MDDEs for the  MDDEs. According to this figure, VB BS MDDE is the best in terms of NSE followed by VB RIS MDDE. Note that all of the MDDEs that use the VB approximate density as the weighting functions have similar values while all other estimators have values quite different from one another. This could be an additional indication of numerical stability of our proposed estimators for this model with non-standard estimation procedures.

\subsection{Longitudinal Poisson Model} 

\noindent The online supplement reports extensive analysis of the MDDEs for the longitudinal Poisson model, the analysis of which constitutes a compelling case as this model that requires a more complicated VB estimation procedure. Our simulation clearly shows that the existing solutions are unsatisfactory regarding the bias and the NSE of the estimators. Only our newly proposed RIS and BS with the VB approximate posterior densities used as the weighting functions together with the Warp3 version of the Bs MDDE seem to stay unbiased and outperformed all other estimators in terms of the NSE.

\section{Conclusions}\label{sec:conclusions}\setcounter{equation}{0}

\noindent We have shown that VB methods, that thus far have been mostly used as a fast alternative to MCMC methods for the estimation of Bayesian models, can be effectively used to estimate the MDD precisely. Our method is applicable to a wide range of models for which a  sample from the posterior is available and VB estimation is feasible. The new MDD estimators are accurate, numerically stable, and simple to compute. They are also at least as precise as the benchmark MDDEs that we considered in our simulations. 

The improvement in the precision comes from the accuracy of the VB approximate posterior density and its optimality in terms of minimizing the reverse KL divergence from the posterior distribution, a property which directly translates into optimality of the MDDE, and results in a minimum variance estimator for the class of RIS MDDEs. We have also presented other desirable properties of our RIS MDDE such as the consistency, unbiasedness, and asymptotic normality of the reciprocal of the RIS MDDE. Finally, it is worth noting that our idea for the MDDE as well as all of the theoretical results for the RIS MDDE do not depend on the mean-field VB factorization as in equation (\ref{eq:vbassumption}) and apply to VB approximate posterior densities free of this assumption.

\bibliographystyle{ba}
\bibliography{sample}

\newpage
\begin{appendix}
\section{Proofs}

\subsection*{Proof of Proposition \ref{cl:dominance}}
Note that if a solution $q^*$ to problem from equation \eqref{eq:minkl} exists, it must satisfy:
\begin{equation*}
\int q^* \left(\theta\right)\ln\left[ \frac{q^*\left(\theta\right)}{p(\theta|\mathbf{y})} \right]d\theta<\infty.
\end{equation*}
Then, consider the following identity:
\begin{multline*}
\int q^* \left(\theta\right)\ln\left[ \frac{q^*\left(\theta\right)}{p(\theta|\mathbf{y})} \right]d\theta = 
\int_{\Omega_p\cap\Omega_q}q^*\left(\theta\right)\ln\left[ \frac{q^*\left(\theta\right)}{p(\theta|\mathbf{y})} \right]d\theta \\
+ \int_{\Omega_p^c\cap\Omega_q}q^*\left(\theta\right)\ln\left[ \frac{q^*\left(\theta\right)}{p(\theta|\mathbf{y})} \right]d\theta 
+ \int_{\Omega_q^c}q^*\left(\theta\right)\ln\left[ \frac{q^*\left(\theta\right)}{p(\theta|\mathbf{y})} \right]d\theta.
\end{multline*}
Note that the first integral on the right hand side is non-negative. We follow the conventions used in the context of the KL divergence, e.g. in \cite{Cover2006}, to set $0\ln\frac{0}{0}=0$, $0\ln\frac{0}{p}=0$, and $q\ln\frac{q}{0}=\infty$. Therefore, the last integral is zero. Now, suppose that $p(\mathbf{y}|\theta) p(\theta)=0$ but $q^*(\theta)\neq0$ over a set with non-zero measure. Then, $q\ln\frac{q}{p}=\infty$ over a set with non-zero measure making the second integral on the right hand side infinite. This is inconsistent with KL minimisation having a solution and, thus, the dominance property must hold .

\subsection*{Proof of Proposition \ref{prop:MDDinv}}

The proposition requires  $p(\mathbf{y})$ to be finite which holds under Assumption \ref{as:densities} as shown by \cite{ChanEisenstat2015}. According to Bayes' theorem: 
$$\frac{p(\theta|\mathbf{y})}{p(\mathbf{y}|\theta)p(\theta)}=\frac{1}{p(\mathbf{y})}.$$
Therefore, we can write:
\begin{equation*}\label{eq:int1}
\begin{split}
& \int_{\Omega_p}\frac{q^{*}(\theta)}{p(\mathbf{y}|\theta)p(\theta)}p(\theta|\mathbf{y})d\theta
= \frac{1}{p(\mathbf{y})}\int_{ \Omega_p}^{}q^{*}(\theta)d\theta \\
&= \frac{1}{p(\mathbf{y})}\left[\int_{ \Omega_q}^{}q^{*}(\theta)d\theta-\int_{ \Omega_q \cap \Omega_p^c}^{}q^{*}(\theta)d\theta+\int_{ \Omega_p \cap \Omega_q^c}q^{*}(\theta)d\theta\right]\\
&= \frac{1}{p(\mathbf{y})}\left[1-\int_{ \Omega_q \cap \Omega_p^c}^{}q^{*}(\theta)d\theta\right]
\end{split}
\end{equation*}
Note that the third integral in the second line of the equation above is zero and the integral in the last line is zero if the property from Proposition \ref{cl:dominance} holds, which concludes the proof.

\subsection*{Proof of Proposition \ref{prop:var}}

The variance of $\frac{q^*}{p}$ can be written as:
\begin{equation*}\label{eq:var}
\begin{split}
Var_p \left(\dfrac{q^*}{p}\right) &= E_p\left(\frac{q^*}{p}\right)^2-\left[E_p\left(\frac{q^*}{p}\right)\right]^2 = E_q\left(\dfrac{q^*}{p}\right)-1\\
&= \int_{\Omega_q}\left[\frac{q^{*}(\theta)}{p(\theta|\mathbf{y})}-1\right]q^{*}(\theta|\mathbf{y})d\theta,
\end{split}
\end{equation*}
where the first element in the difference from the first line of the equation above above is:
\begin{align*}
E_p \left(\frac{q}{p}\right)^2 &= \int_{\Omega_p} \left[\frac{q^{*}(\theta)}{p(\theta|\mathbf{y})}\right]^2p(\theta|\mathbf{y})d\theta \\
&= \int_{\Omega_p} \left[\frac{q^{*}(\theta)}{p(\theta|\mathbf{y})}\right]q^{*}(\theta)d\theta\\
&= \int_{\Omega_q} \left[\frac{q^{*}(\theta)}{p(\theta|\mathbf{y})}\right]q^{*}(\theta)d\theta = E_q\left[\frac{q^*}{p}\right]
\end{align*}
and 
\begin{equation*}
E_p \left(\frac{q}{p}\right) =\int_{\Omega_p}\left[\frac{q^{*}(\theta)}{p(\theta|\mathbf{y})}\right]p(\theta|\mathbf{y})d\theta 
=\int_{\Omega_p}q^{*}(\theta)d\theta=\int_{\Omega_q}q^{*}(\theta)d\theta=1
\end{equation*}
Some of these equalities hold only if $p(\mathbf{y}|\theta) p(\theta)=0\Rightarrow q^{*}(\theta)=0$ is true almost everywhere.

\subsection*{Proof of Proposition \ref{prop:pris}}

Given the assumptions and Proposition \ref{prop:var}, part (i) can be proved by direct application of an ergodic central limit theorem \citep[see e.g.][Theorem 5]{Tierney1994}. For the formula of the variance of the asymptotic normal distribution, see the discussions in \cite{Geweke1992}, \cite{Chib1995}, or \cite{Fruhwirth-Schnatter2006}. The latter reference provides an estimator for $\rho_h(0)$.

\noindent To prove part (ii)  write $\hat p_{RIS.VB}(\mathbf{y})^{-1}$ as the average of $S$ values of the random variable $d_s$, that is:
\begin{equation*}
\hat p_{RIS.VB}(\mathbf{y})^{-1} = \frac{1}{S}\sum_{s=1}^{S}d_s.
\end{equation*}
Then, we have:
\begin{equation*}
E[d_s] = \int \frac{q^*\left(\theta^{(s)}\right)}{p\left(\mathbf{y}|\theta^{(s)}\right)p\left(\theta^{(s)}\right)}p\left(\theta^{(s)}|\mathbf{y}\right)d\theta^{(s)}
= p(\mathbf{y})^{-1} \int q^*\left(\theta^{(s)}\right) d\theta^{(s)} = p(\mathbf{y})^{-1}<\infty,
\end{equation*}
where the last inequality comes from the fact that $p(\mathbf{y})>0$. Then
\begin{equation*}
E\left[ \hat p_{RIS.VB}(\mathbf{y})^{-1} \right] = E\left[ \frac{1}{S}\sum_{s=1}^{S}d_s \right] =  \frac{1}{S}\sum_{s=1}^{S}E\left[d_s \right] = p(\mathbf{y})^{-1}.
\end{equation*}

\subsection*{Proof of Proposition \ref{prop:minvar}}

VB is  commonly defined based on KL divergence but it is perfectly possible to define a variational approximation based on other well-defined notions of divergence \citep[see e.g.][]{Zhang2017}. One such possibility is the Pearson $\chi^2$ divergence criterion which is a natural upper bound for KL divergence \citep[see e.g.][]{Wang2016}. In this case a variational estimator can be defined as:
\begin{equation*}\label{eq:minchi}
q^*(\theta)=\underset{q }{\argmin} \int\left[\frac{q(\theta)}{p(\theta|\mathbf{y})}-1\right]q(\theta)d\theta
\end{equation*}

\smallskip\noindent First note that if a solution $q^*$ to the problem above exists, it must satisfy:
\begin{equation*}
\int_{\Omega_q}\left[\frac{q^{*}(\theta)}{p(\theta|\mathbf{y})}\right]q^{*}(\theta|\mathbf{y})d\theta<\infty
\end{equation*}
as we can write that:
\begin{align*}
\int&\left[\frac{q^{*}(\theta)}{p(\theta|\mathbf{y})}\right]q^{*}(\theta)d\theta 
=\int_{ \Omega_q\cap \Omega_p} \left[\frac{q^{*}(\theta)}{p(\theta|\mathbf{y})}q^{*}(\theta)\right]d\theta \\
&+ \int_{ \Omega_q \cap \Omega_p^c} \left[\frac{q^{*}(\theta)}{p(\theta|\mathbf{y})}q^{*}(\theta)\right]d\theta 
+ \int_{ \Omega_q^c} \left[\frac{q^{*}(\theta)}{p(\theta|\mathbf{y})}q^{*}(\theta)\right]d\theta,
\end{align*}
and, if a solution $q^*$ exists, the right hand side must be finite as well. The last integral is zero and therefore the sum of the first two integrals on the right hand side, which is equal to  $\int_{\Omega_q}\left[\frac{q^{*}(\theta)}{p(\theta|\mathbf{y})}\right]q^{*}(\theta)d\theta$ must be finite. 

Using proposition \ref{prop:var}, $q^*(\theta)$ is shown to be the candidate density that minimizes the variance of the reciprocal importance sampling estimator and, therefore, it is an optimal candidate.

\section{Details of Variational Bayes Estimation for Two Models}

\subsection{Vector Autoregressive Model}

\paragraph{Normal-Wishart Conjugate Prior} 
The VB lower bound for the $\ln MDD$ for the VAR model with normal-Wishart conjugate prior is not needed for estimation as the analytical formulae for the hyper-parameters of the VB posterior distribution are given in a close-form in the paper. We report the value below because we use it in the numerical exercise for the MDDs.
\begin{align*}
\ln MDD_{VBLB} = &- \frac{NT}{2}\ln\pi +\frac{NK}{2}\ln(2e) +\frac{N}{2}\left( (T+\underline{\nu})\ln(T+\underline{\nu}) - \overline{\nu}^{*}\ln\overline{\nu}^{*} \right)\\
& + \ln\Gamma_{N}\left( \frac{\overline{\nu}^{*}}{2} \right) - \ln\Gamma_{N}\left( \frac{\underline{\nu}}{2} \right) 
+ \frac{N}{2}\left( \ln\left| \overline{V}^{*}\right| - \ln\left| \underline{V}\right| \right)\\
& +\frac{1}{2} \left( \left(T+\underline{\nu}\right)\ln\left| \overline{S}^{*-1}\right| - \underline{\nu}\ln\left| \underline{S}^{-1} \right| \right)
\end{align*}

\paragraph{Normal-Wishart Independent Prior}
The optimal VB approximate posterior distributions are the multivariate normal distribution for $\alpha$ and the Wishart distribution for $\Sigma^{-1}$:
\begin{equation*}
q_{\alpha}^{*}\left(\alpha\right) =\mathcal{N}\left( \overline{\alpha}^{*}, \overline{V}^{*} \right),\text{ and }
q_{\Sigma}^{*}\left(\Sigma^{-1}\right) =\mathcal{W}\left( \overline{S}^{*-1}, \overline{\nu}^{*} \right),
\end{equation*}
where the optimal values of the parameters determining the optimal VB approximate posterior distributions are obtained using an iterative procedure. Initialize by setting starting values for $\overline{S}$ and $\overline{\nu}$, and iterate:
\begin{align*}
\overline{\alpha} & \leftarrow \overline{V}\left( \underline{V}^{-1}\underline{\alpha} + \overline{\nu}\sum_{t=1}^{T}(x_t^{'}\otimes I_N)\overline{S}y_t \right),\\
\overline{V} & \leftarrow \left( \underline{V}^{-1} + \overline{\nu}\sum_{t=1}^{T}(x_t^{'}\otimes I_N)\overline{S}(x_t^{'}\otimes I_N)^{'} \right)^{-1},\\
\overline{S} &\leftarrow \underline{S} + \sum_{t=1}^{T}\left[\left( y_t - (x_t^{'}\otimes I_N)\overline{\alpha} \right)\left( y_t - (x_t^{'}\otimes I_N)\overline{\alpha} \right)^{'} \right. \\
&\qquad \left. + (x_t^{'}\otimes I_N)\overline{V}(x_t^{'}\otimes I_N)^{'} \right],\\
\overline{\nu} &\leftarrow T + \underline{\nu},
\end{align*}
until the increase in the corresponding $\ln MDD_{VBLB}$ is negligible where the latter value is given by: 
\begin{align*}
\ln MDD_{VBLB} = &\frac{k}{2} - \frac{NT}{2}\ln\pi + \ln\Gamma_{N}\left( \frac{\overline{\nu}^{*}}{2} \right) - \ln\Gamma_{N}\left( \frac{\underline{\nu}}{2} \right) + \frac{1}{2}\left( \ln\left| \overline{V}^{*}\right| - \ln\left| \underline{\underline{V}}\right| \right)\\
&+\frac{1}{2} \left( \overline{\nu}^{*}\ln\left| \overline{S}^{*-1}\right| - \underline{\nu}\ln\left| \underline{S}^{-1} \right| \right) - \frac{1}{2}\text{tr}\left\{ \underline{\underline{V}}^{-1} \left[\left(\overline{\alpha}^{*} - \underline{\alpha}  \right)\left(\overline{\alpha}^{*} - \underline{\alpha}  \right)^{'} + \overline{V}^{*}  \right] \right\}.
\end{align*}

\subsection{Stochastic Frontier Model} 

\paragraph{Exponential inefficiency}
The optimal VB approximate posterior distributions turn out to be:
\begin{align*}
q_{\beta}^{*}(\beta)&= \mathcal{N}\left(\overline{\beta}^{*},\overline{V}_{\beta}^{*} \right), \quad
q_{\sigma^{-2}}^{*}\left(\sigma^{-2}\right)= \mathcal{G}\left(\overline{A}_{\sigma}^{*}, \overline{B}_{\sigma}^{*}\right), \\
q_{\lambda}^{*}(\lambda)&= \mathcal{G}\left(\overline{A}_{\lambda}^{*}, \overline{B}_{\lambda}^{*} \right), \quad
q_u^{*}\left(u_i\right)= \mathcal{TN}\left(\overline{\mu}_i^{*},\overline{\upsilon}^{2*} \right),
\end{align*}
where $\mathcal{TN}(\cdot ,\cdot )$ denotes the normal density function truncated to positive values of $u_i$.

The optimal hyper-parameters characterizing these VB approximate posterior distributions are computed by the coordinate ascent algorithm over the following iterations:
\begin{align*}
\overline{\beta} &\leftarrow \overline{V}_{\beta} \left(\underline{V}_{\beta}^{-1}\underline{\beta} + \overline{A}_{\sigma}\overline{B}_{\sigma}^{-1} x'\left(y \pm \overline{u} \otimes \imath_T\right) \right) ,\\
\overline{V}_{\beta} &\leftarrow \left( \underline{V}_{\beta}^{-1} + \overline{A}_{\sigma}\overline{B}_{\sigma}^{-1} x'x \right)^{-1}, \\
\overline{A}_{\sigma} &\leftarrow \underline{A}_{\sigma} + 0.5NT, \\
\overline{B}_{\sigma} &\leftarrow \underline{B}_{\sigma}+\frac{1}{2}\left( \sum_{i=1}^{N}\sum_{t=1}^{T}\left[ \left( y_{it}-x_{it}\overline{\beta}\pm \overline{u}_{i} \right)^{2}+\overline{Var}\left[ u_{i} \right] \right] +\text{tr}\left(x'x \overline{V}_{\beta} \right) \right), \\
A_{\lambda} &\leftarrow \underline{A}_{\lambda} + N, \\
B_{\lambda} &\leftarrow \underline{B_{\lambda}} + \sum_{i=1}^{N}\overline{u}_i, \\
\overline{\mu}_{i} &\leftarrow \frac{1}{T}\left( - \overline{A}_{\lambda}\overline{B}_{\lambda}^{-1}\overline{A}_{\sigma}^{-1}\overline{B}_{\sigma} \pm \sum_{t=1}^{T}\left(x_{it}\overline{\beta} - y_{it} \right)  \right), \\
\overline{\upsilon}^{2} &\leftarrow T^{-1}\overline{A}_{\sigma}^{-1}\overline{B}_{\sigma},
\end{align*}
where $x$ is an $NT\times k$ matrix with its $(i,t)$th row set to $x_{it}$,  $\imath_{T}$ is a $T$-vector of ones, and $\overline{u}=(\overline{u}_1,\dots,\overline{u}_N)'$. The quantities $\overline{u}_i = \overline{\mu}_i + \overline{\upsilon} m\left( \overline{\mu}_i / \overline{\upsilon} \right)$, and $$\overline{Var}\left[u_i\right] = \overline{\upsilon}^2 \left\{ 1-m\left( \overline{\mu}_i / \overline{\upsilon} \right)\left[ m\left( \overline{\mu}_i / \overline{\upsilon} \right)+ \overline{\mu}_i / \overline{\upsilon} \right]\right\}$$ are the corresponding moments of the truncated normal distribution for $u_i$. They can be computed using $m(\cdot)=\phi(\cdot)/\Phi(\cdot)$, where $\phi(\cdot)$ and $\Phi(\cdot)$ are the pdf and the cdf of the standard normal distribution, respectively. We iterate the hyper-parameters of the approximate posterior distributions until the increment in the corresponding value of $\ln MDD_{VBLB}$ for the model is negligible. An~analytical formula for the latter value can be found in the Appendix.

\begin{multline*}
\ln MDD_{VBLB} = \frac{1}{2}((-NT+N)\ln 2\pi +N+k)
+\ln\left(\frac{\Gamma \left(\overline{A}_{\sigma}^{*} \right) \Gamma \left(\overline{A}_{\lambda}^{*} \right)\underline{B}_{\sigma }^{\underline{A}_{\sigma}}\underline{B}_{\lambda}^{\underline{A}_{\lambda}}}{\Gamma \left(\underline{A}_{\sigma} \right)\Gamma \left( \underline{A}_{\lambda} \right)\overline{B}_{\lambda }^{*\overline{A}_{\lambda}^{*}}\overline{B}_{\sigma}^{*\overline{A}_{\sigma}^{*}}} \right)\\
+\frac{1}{2}\ln \left( \frac{|\overline{V}_{\beta}^{*}|}{|\underline{V}_{\beta}| }\right) 
-\frac{1}{2}\left(\overline{\beta}^{*}-\underline{\beta}\right)'\underline{V}_{\beta}^{-1}\left(\overline{\beta}^{*}-\underline{\beta}\right)\\
+\text{tr}(\underline{V}_{\beta}^{-1}\overline{V}_{\beta}^{*}) 
+N\ln\overline{\upsilon}^{*}
 + \sum_{i=1}^{N}\left( \ln\Phi \left( \frac{\overline{\mu}_i^{*} }{ \overline{\upsilon}^{*}} \right)-\frac{1}{2}\frac{\overline{\mu}_i^{*} }{ \overline{\upsilon}^{*}}m\left(\frac{\overline{\mu}_i^{*} }{ \overline{\upsilon}^{*}} \right)\right)
\end{multline*}

\paragraph{Gamma inefficiency}
By applying a suitable factorization, it can be shown that the VB optimal densities are given by:
\begin{equation*}
q\left(\beta, \sigma, \lambda, \theta, u\right) = 
q_{\beta}(\beta)q_{\sigma}\left(\sigma^{-2}\right)q_{\lambda}(\lambda)q_{\theta}(\theta)q_u\left(u\right),
\end{equation*}
The first three of these distributions have functional forms of standard distributions:
\begin{equation*}
q_{\beta}^{*}(\beta) = \mathcal{N}\left(\overline{\beta}^{*},\overline{V}_{\beta}^{*} \right),\quad
q_{\sigma}^{*}\left(\sigma^{-2}\right) = \mathcal{G}\left(\overline{A}_{\sigma}^{*},\overline{B}_{\sigma}^{*} \right),\quad
q_{\lambda}^{*}(\lambda) = \mathcal{G}\left(\overline{A}_{\lambda}^{*},\overline{B}_{\lambda}^{*} \right),
\end{equation*}
while the approximate posterior distribution for $u$ has a nonstandard form of:
\begin{equation*}
q_u^{*}\left(u_i\right) = \frac{u_{i}^{\overline{\theta}^{*} - 1} \exp\left\{ -\frac{\overline{\upsilon}^{*2}}{2}u_{i}^{2} - \overline{\mu}_i^{*} u_i \right\}}{\overline{\upsilon}^{*-\overline{\theta}^{*}} \exp\left\{ \overline{\mu}_{i}^{*2}/4\overline{\upsilon}^{*2} \right\} \Gamma\left(\overline{\theta}^{*}\right) D_{-\overline{\theta}^{*}}\left(\overline{\mu}_i^{*}/\overline{\upsilon}^{*}\right)},
\end{equation*}
where $D_{-\overline{\theta}}(\cdot)$ denotes the parabolic cylinder function. The results in \citeauthor{beckers1987tractable}~(\citeyear{beckers1987tractable}, pp.~30) have been used to find the normalizing constant of the distribution above. Finally, the approximate posterior distribution for $\theta$ is known up to its normalizing constant, $c_{\theta}$:
\begin{equation*}
q_{\theta}^{*}(\theta) =  c_{\theta}^{-1} \theta^{-\underline{A}_{\theta}+1} \exp\left\{ \left( (N+1)\overline{\ln\lambda} + \ln\underline{B}_{\lambda} + \sum_{i=1}^{N}\overline{\ln u_i} \right) \theta  - \underline{B}_{\theta} \theta^{-1} - (N+1)\ln\Gamma (\theta)\right\},
\end{equation*}

The hyper-parameters of the VB approximate posterior distributions for this model are determined by the coordinate ascent algorithm using the following iterations:
\begin{align*}
\overline{\beta} &\leftarrow \overline{V}_{\beta}\left(\underline{V}_{\beta}^{-1}\underline{\beta} + \overline{A}_{\sigma}\overline{B}_{\sigma}^{-1} x'\left(y \pm \overline{u} \otimes \imath_T\right) \right)\\
\overline{V}_{\beta} &\leftarrow \left(\underline{V}_{\beta}^{-1} + \overline{A}_{\sigma}\overline{B}_{\sigma}^{-1} x'x \right)^{-1}\\
\overline{A}_{\sigma} &\leftarrow \underline{A}_{\sigma}+\frac{NT}{2}\\
\overline{B}_{\sigma} &\leftarrow \underline{B}_{\sigma}+\frac{1}{2}\left( \sum_{i=1}^{N}\sum_{t=1}^{T}\left( \left(y_{it} - x_{it}\overline{\beta} \pm \overline{u}_i \right)^2 + Var\left[u_i\right]\right) + \text{tr}\left(x'x \overline{V}_{\beta}\right)\right)\\
\overline{A}_{\lambda} &\leftarrow (N+1)\overline{\theta}\\
\overline{B}_{\lambda} &\leftarrow \sum_{i=1}^{N} \underline{B}_{\lambda} + \overline{u}_i \\
\overline{\upsilon}^{2} &\leftarrow T\overline{A}_{\sigma}\overline{B}_{\sigma}^{-1}\\
\overline{\mu}_{i} &\leftarrow \left( - \overline{A}_{\lambda}\overline{B}_{\lambda}^{-1} \pm\sum_{t=1}^{T} \overline{A}_{\sigma}\overline{B}_{\sigma}^{-1} \left(y_{it} - x_{it} \overline{\beta}\right) \right),
\end{align*}   
where $ \overline{\theta}= \int_{0}^{\infty}\theta q_{\theta}^{*}(\theta)d\theta$, and $Var\left[u_i\right] = \overline{u_{i}^{2}} - \overline{u}_{i}^{2}$, while the appropriate moments of $u_i$ are computed as 
$\overline{u_{i}}^{m}=\frac{\Gamma\left(\overline{\theta}+ m\right) D_{-\overline{\theta} - m}\left( \mu_{i}/\upsilon \right) }{\upsilon^{m} \Gamma\left(\overline{\theta}\right) D_{ -\overline{\theta}}\left(\mu_i/\upsilon\right)}$.
The optimal densities are obtained by iterating the quantities  above until increment of $\ln MDD_{VBLB}$ is negligible.

\begin{multline*}
\ln MDD_{VBLB.SF.G} = 
\frac{-NT\ln 2\pi +k}{2}
+ \ln\left(\frac{\Gamma\left(\overline{A}_{\sigma}^{*} \right)\Gamma\left(\overline{A}^{*}_{\lambda}\right)\underline{B}_{\sigma}^{\underline{A}_{\sigma}}\underline{B}_{\lambda}^{\overline{\theta}^{*}} \underline{B}_{\theta}^{\underline{A}_{\theta}}}{\Gamma \left(\underline{A}_{\sigma}\right)\Gamma\left(\underline{A}_{\theta}\right)\overline{B}_{\sigma}^{*\overline{A}_{\sigma}^{*}}\overline{B}_{\lambda}^{*\overline{A}_{\lambda}^{*}}} \right)
\\
- \frac{1}{2}\left(\overline{\beta}^{*} - \underline{\beta}\right)' \underline{V}_{\beta}^{-1}\left(\overline{\beta}^{*} - \underline{\beta}\right)
 - \frac{1}{2}\text{tr}\left(\underline{V}_{\beta}^{-1} \overline{V}_{\beta}^{*}\right)
 \\+ \frac{1}{2}\ln\left(\frac{|\overline{V}_{\beta}^{*}|}{|\underline{V}_{\beta}|} \right)
+ \left(\overline{\theta}^{*} - 1\right) \sum_{i=1}^{N} \overline{\ln u_i} 
- \sum_{i=1}^{N}\overline{\ln q\left(u_i \right)}
\\- (N+1)\overline{\ln \Gamma (\theta)}
- \left(\underline{A}_{\theta} - 1 \right)\overline{\ln \theta}
- \underline{B}_{\theta}\overline{\theta^{-1}}-\overline{\ln q(\theta)}
\end{multline*}
where:
$$\overline{\ln\theta} = \int\limits_{0}^{\infty}\ln\theta q_{\theta}^{*}(\theta)d\theta,\quad 
\overline{\ln q(\theta)} = \int\limits_{0}^{\infty}\ln q_{\theta}(\theta)q_{\theta}^{*}(\theta)d\theta, \quad\overline{\ln u_i} = \int\limits_{0}^{\infty} \ln u_i q_{u}(u_i)du_i$$


\section{Numerical Accuracy of the MDD Estimators for Longitudinal Poisson Model}

\noindent We chose the example of the longitudinal Poisson model (LPM) to illustrate the performance of our MDDEs for a model that requires a more complicated VB estimation procedure. The LPM was also considered in other works proposing new MDDEs such as those by \cite{Chib1998}, \cite{ChibJeliazkov2001}, and \cite{Perrakis2014}. For the sake of comparison, we adopt similar modeling assumptions including conditional independence for the count data given the explanatory variables:
\begin{equation*}\label{eq:lpm}
y_{it} \sim \mathcal{POISSON}\left\{ \exp\left( a_{it} + x_{it}'\beta + z_{it}'u_i \right)\right\},
\end{equation*}
for $i=1,\dots,N$ and $t=1,\dots,T$, where $x_{it}$, a $k$-vector, and $z_{it}$, an $m$-vector, are the regressors and $u_i$ are $m$-vectors of normally distributed latent variables, $u_i\sim\mathcal{N}(\mu, \Sigma)$. The vectors $\beta$, $\mu$, and the matrix $\Sigma$ are the parameters of the model.

\subsection{Prior and posterior analysis with VB estimation}

Following the papers mentioned above, we assume the following prior distributions for the parameters:
\begin{equation*}\label{eq:lpm-prior}
\beta\sim\mathcal{N}\left( \underline{\beta}, \underline{V}_{\beta} \right), \quad \mu\sim\mathcal{N}\left( \underline{\mu}, \underline{V}_{\mu} \right),  \quad \Sigma^{-1}\sim\mathcal{W}\left( \underline{S}^{-1}, \underline{\nu} \right),
\end{equation*}
and that $a_{it}$ is the offset that is equal to $\ln(8)$ when $t=0$ and $\ln(2)$ otherwise \citep[see e.g.][]{Perrakis2014}. Bayesian estimation of the parameters of the model is nonstandard as only $\mu$ and $\Sigma^{-1}$ have multivariate normal and Wishart full conditional posterior densities respectively, while the conditional densities for $\beta$ and $u_i$ are of unknown parametric form. We utilize WinBUGS to sample from the exact posterior distribution of the parameters of this model.

To present the estimation procedure using mean-field VB, we define a vector $u=\VEC\left[(u_1, \dots, u_N)'\right]$, where the operator $\VEC$ vectorizes column-wise its argument, an $NT$-vector $\mathbf{y}$ collecting $y_{it}$s, an $NT\times k$ matrix $X$ containing vectors $x_{it}$, $T\times m$ matrices $z_i$ with rows being equal to $z_{it}$, an $NT\times Nm$ block-diagonal matrix $Z$ with matrices $z_i$ on its main diagonal, and a $NT\times(k+Nm)$ matrix $C= \begin{bmatrix}X&Z\end{bmatrix}$. Finally, collect all $a_{it}$s in an $NT$-vector $a$, and define a matrix $\Gamma = (\beta', u')$, and the vector of the prior means of $\Gamma$ as $\underline{\Gamma} = (\underline{\beta}',\underline{\mu},\dots,\underline{\mu})$. 

We assume the following factorization of the approximate posterior distribution:
\begin{equation*}\label{eq:mlp-factorization}
q\left(\Gamma,\mu,\Sigma^{-1}\right) = q_{\Gamma}(\Gamma) q_{\mu}(\mu) q_{\Sigma}\left(\Sigma^{-1}\right).
\end{equation*}
The assumption above does not lead to a closed form solution for $q_{\Gamma}^{*}(\Gamma)$. The solution to this problem was proposed by \cite{Luts2015} and assumes that this distribution is well-approximated by a multivariate normal distribution and its parameters computed using an extension of mean-field VB by \cite{Knowles2011} and known as \emph{non-conjugate variational message passing}. The model of \cite{Luts2015} is somewhat different from ours but the same strategy can be applied. The individual approximate posterior distributions are:
\begin{align*}
q_{\Gamma}^{*}(\Gamma) &= \mathcal{N}\left( \overline{\Gamma}^{*}, \overline{V}_{\Gamma}^{*} \right), \\
q_{\mu}^{*}(\mu) &= \mathcal{N}\left( \overline{\mu}^{*}, \overline{V}_{\mu}^{*} \right), \\
q_{\Sigma}^{*}\left(\Sigma^{-1}\right) &= \mathcal{W}\left( \overline{S}^{*}, \overline{\nu}^{*} \right).
\end{align*}
The optimal values of the parameters determining these distribution are obtained by iterating:
\begin{align*}
\overline{V}_{\Gamma} &\leftarrow C' \diag\left[ a + \exp\left\{ C\overline{\Gamma} \right\} + \frac{1}{2}\diag\left( C\overline{V}_{\Gamma}C' \right) \right]C  + \begin{bmatrix} \underline{V}_{\beta}^{-1} & \mathbf{0} \\ \mathbf{0} & \overline{\nu}I_N\otimes \overline{S} \end{bmatrix}\\
\overline{\Gamma} &\leftarrow \overline{\Gamma} + \overline{V}_{\Gamma}\left[ C'\left( \mathbf{y} - \exp\left\{ a + C\underline{\Gamma} + \frac{1}{2}\diag\left( C\overline{V}_{\Gamma}C' \right)  \right\} \right)  -  \begin{bmatrix} \underline{V}_{\beta}^{-1} & \mathbf{0} \\ \mathbf{0} & \overline{\nu}I_N\otimes \overline{S} \end{bmatrix} \left( \overline{\Gamma} - \begin{bmatrix} \underline{\beta} \\ \imath_N \otimes \overline{\mu}\end{bmatrix} \right) \right]\\
\overline{V}_{\mu} &\leftarrow \left(\underline{V}_{\mu}^{-1} + N\overline{\nu}\overline{S}\right)^{-1}\\
\overline{\mu} &\leftarrow \overline{\nu}\overline{V}_{\mu}\overline{S} \left[ \sum_{i=1}^{N}\left( \overline{u}_i - \underline{\mu} \right) \right]\\
\overline{S} &\leftarrow \left[ \underline{S} + N\overline{V}_{\mu} + \sum_{i=1}^{N}\left[ \left( \overline{u}_i - \overline{\mu} \right)\left( \overline{u}_i - \overline{\mu} \right)' + \overline{V}_{u_i} \right] \right]^{-1} \\
\overline{\nu} &\leftarrow \underline{\nu} + N
\end{align*}
where $\overline{u_i}$, and $\overline{V}_{u_i}$, are relevant sub-matrices of $\overline{\Gamma}$ and $\overline{V}_{\Gamma}$ respectively, until the increment in $\ln MDD$ is negligible. An analytical formula for the latter value is given in the Appendix.

\begin{multline*}\label{eq:lmp-vblb}
\ln MDD_{VBLB.LMP} = \imath_{NT}' \exp\left\{ a + C \overline{\Gamma}  + \frac{1}{2}\diag\left( C\overline{V}_{\Gamma}C' \right) \right\} + y'\left( a + C\overline{\Gamma} \right) -\imath_{NT}' \ln y! \\
- \frac{1}{2} \tr\left[ \underline{V}_{\beta}^{-1} \left( \overline{\beta} - \underline{\beta} \right)\left( \overline{\beta} - \underline{\beta} \right)' + \overline{V}_{\beta} \right] - \frac{1}{2} \tr\left[ \underline{V}_{\mu}^{-1} \left( \overline{\mu} - \underline{\mu} \right)\left( \overline{\mu} - \underline{\mu} \right)' + \overline{V}_{\mu} \right] + \frac{1}{2}\ln\frac{|\overline{V}_{\Gamma}|}{|\underline{V}_{\beta}|} + \frac{1}{2}\ln\frac{|\overline{V}_{\mu}|}{|\underline{V}_{\mu}|} \\
+ \frac{\underline{\nu}}{2}\ln |\underline{S}|  - \frac{\overline{\nu}}{2}\ln |\overline{S}|  
+ \ln\Gamma_m\left( \frac{\overline{\nu}}{2} \right) - \ln\Gamma_m\left( \frac{\underline{\nu}}{2} \right) + \frac{1}{2}\left(k + m + Nm + Nm\ln2 \right)
\end{multline*}

\subsection{Comparison of the MDD estimators}

\begin{table}[t!]
\raggedright
\caption{Numerical accuracy of the MDDE for Longitudinal Poisson Model}
\begin{center}\small
\begin{tabular}{lccccc}
\toprule
\multicolumn{6}{c}{\textbf{Longitudinal Poisson Model}}\\
\midrule
&\multicolumn{2}{l}{\textit{Benchmark values}}&\multicolumn{3}{l}{\textit{Benchmark MDDEs}}\\
 & VB$_{LB}$ & VB$_{UB}$& Chib, Jeliazkov & Chib et al.  & IS PMPD  \\
$\ln\hat{p}(\mathbf{y})$& -917.2 & -911.2& \textbf{-915.23}$^{*}$ & \textbf{-915.49}$^{*}$& \textbf{-914.992}$^{*}$\\
SE$_{BM}$&	&	0.385 &&& 0.035$^{*}$ \\
NSE &&0.119&& &\\[1ex]
\midrule
&\multicolumn{4}{l}{\textit{RIS MDDEs}}&\\
$h(\theta)$& VB & Geweke & Newton, Raftery&  Lenk&\\
$\ln\hat{p}(\mathbf{y})$& \textbf{-915.3} &  -917.3 & -794.6 & -917.4&\\
SE$_{BM}$&	0.019 & 0.025 & 4.573&	7.694&\\
NSE&	0.031 & 2.181 & 3.019&	1.979&\\
\% within & 100 & 66 & 0  & 59& \\[1ex]
\midrule
&\multicolumn{3}{l}{\textit{BS MDDEs}} &&\textit{IS MDDE}\\
$f(\theta)$ & VB & Normal & Warp3  && VB \\
$\ln\hat{p}(\mathbf{y})$&\textbf{-915.3} & \textbf{-915.5} & \textbf{-915.3} && \textbf{-915.4} \\
SE$_{BM}$&	 0.017 & 0.001 & &&0.091\\
NSE&	 0.021 & 0.006 & 0.020&&0.202\\
\% within &   100 & 100 & 100 && 100\\[1ex]
\bottomrule
\end{tabular}
\end{center}\small
\label{tab:nselpm}
Note: The standard errors computed by the batch means method (SE$_{BM}$) are based on 30 batches each utilizing 1000 draws from the posterior distribution as in \cite{Perrakis2014}. Values denoted by $^{*}$ are taken from \cite{Perrakis2014}.
\end{table}

To illustrate the performance of our new estimators for the longitudinal Poisson model we consider the data set from \cite{Diggle1995}, consisting of seizure counts $y_{it}$ from a group of epileptics ($i=1,2,\dots,59$) monitored initially over an 8-week baseline period ($t=0$) and then over four subsequent 2-week periods ($t=1,2,3,4$). Each patient is randomly assigned either a placebo or the drug progabide after the baseline period. 

\begin{figure}[H]
\raggedright
\begin{center}
\caption{Comparison of simulation outcomes for the best MDDEs for Longitudinal Poisson model}
\includegraphics[trim=1cm 1.5cm 1cm 1cm, scale=0.32]{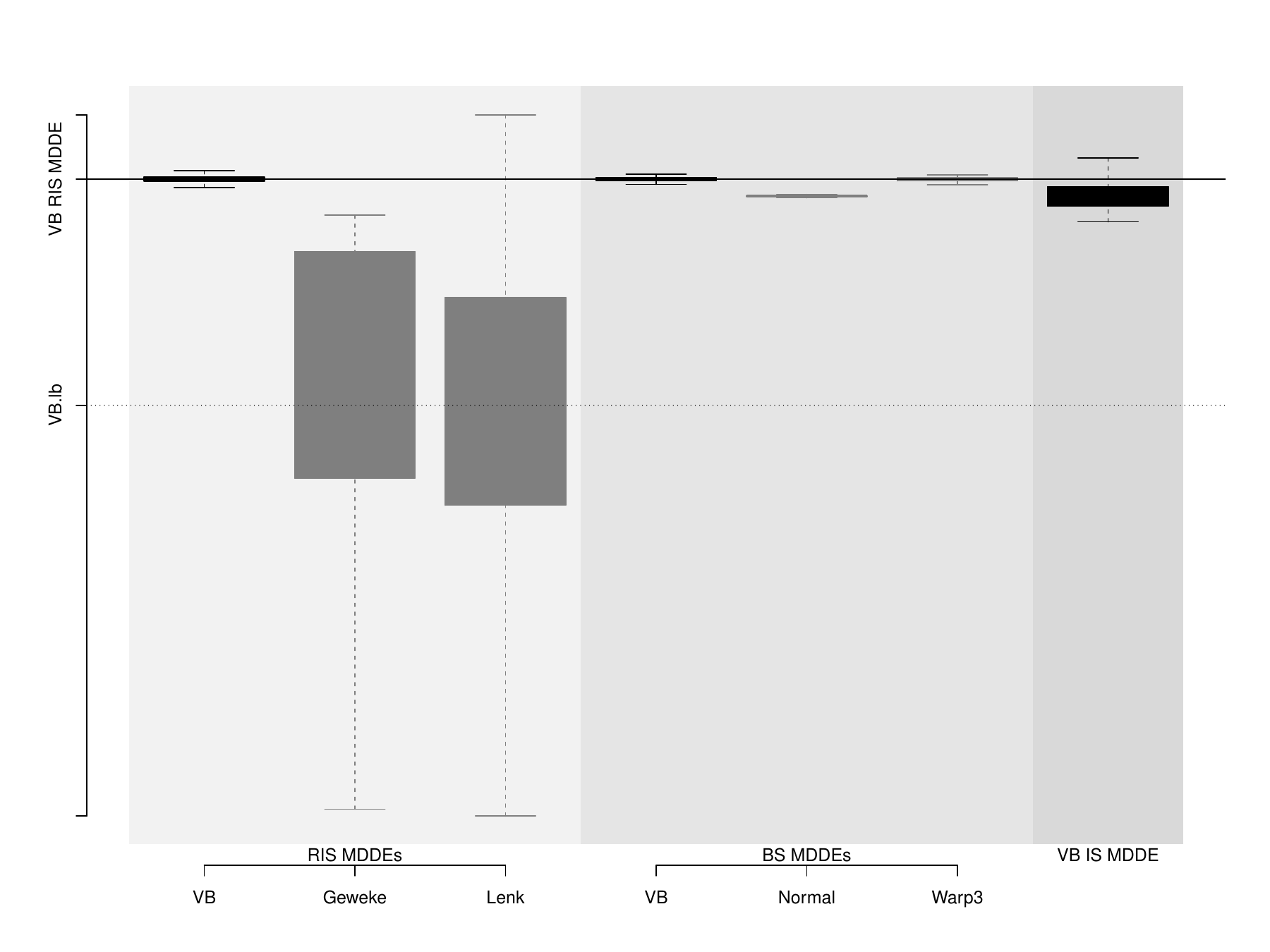}
\label{fig:MDD-lpm}
\end{center}
\small
Note: The figure reports box plots of the natural logarithms of the MDDEs for 100 independent repetitions of the estimation process. Black box of MDDEs with VB approximate posterior distribution as a weighting function are black. The horizontal line in plot corresponds to the value of RIS VB MDDE. The RIS MDDEs by Newton \& Raftery is omitted from the plot.
\end{figure}

We use the same data and estimate the same model as \cite{Perrakis2014}. In Table \ref{tab:nselpm} we compare the numerical precision of our estimators with the results reported by \cite{Perrakis2014} in terms of the standard errors computed by batch means, $SE_{BM}$. Both of our new estimators have lower $SE_{BM}$ than the IS MDDEs estimators in \cite{Perrakis2014}. In particular, both of them improve the precision of the MDD estimation when compared with the PMPD IS MDDE using the likelihood function with analytically integrated out latent factors $u_i$. Our two new estimators also outperform other MDDEs in terms of the NSE except for the Normal and Warp3 BS MDDEs. The latter has the NSE nearly equal to the NSE of our VB BS MDDE, the former, however, is slightly downward biased which is well illustrated in Figure \ref{fig:MDD-lpm}. This figure displays the box plots of the logarithm of the simulated MDDEs for 100 repetitions of the estimation for the MDDEs that we have estimated. According to Figure \ref{fig:MDD-lpm}, VB BS MDDE and VB RIS MDDE are precise and their values are close to each other. Normal BS MDDE  

\noindent has a very low NSE but is biased while VB IS MDDE seems more dispersed. Finally, our IS MDDE with the VB approximate posterior density as the weighting function performs reasonably where its precision is an order of magnitude smaller than all of the alternatives.
\end{appendix}

\end{document}